\documentclass[aps,prd,11pt, preprintnumbers,numbers,sort&compress,nofootinbib]{revtex4}

\usepackage{amsmath}
\usepackage{graphicx}
\usepackage{amssymb}
\usepackage{hyperref}

\usepackage{axodraw4j}
\usepackage{pstricks}
\usepackage{color}

\newcommand{\junk}[1]{}

\begin{document}

\title{Vector model in various dimensions} 
\author{Mikhail Goykhman}
\email{michael.goykhman@mail.huji.ac.il}
\author{Michael Smolkin}
\email{michael.smolkin@mail.huji.ac.il}
\affiliation {The Racah Institute of Physics, The Hebrew University of Jerusalem, Jerusalem 91904, Israel}


\begin{abstract}
We study behaviour of the critical $O(N)$ vector model with quartic interaction in $2\leq d \leq 6$ dimensions to the next-to-leading order in the large-$N$ expansion. We derive and perform consistency checks that provide an evidence for the existence of a non-trivial fixed point and explore the corresponding CFT. In particular, we use conformal techniques to calculate the multi-loop diagrams up to and including 4 loops in general dimension. These results are used to calculate a new CFT data associated with the three-point function of the Hubbard-Stratonovich field. In $6-\epsilon$ dimensions our results match their counterparts obtained within a proposed alternative description of the model in terms of $N+1$ massless scalars with cubic interactions. In $d=3$ we find that the OPE coefficient vanishes up to $\mathcal{O}(1/N^{3/2})$ order.
\end{abstract}
 

\maketitle
\section{Introduction}

In the past elementary particle theory was based on the assumption that nature must be described by a renormalizable quantum field theory. However, a dramatic progress in the realm of critical phenomena revealed that any non-renormalizable theory will look as if it were renormalizable at sufficiently low energies. From this perspective renormalizable theories correspond to a subset of RG flow trajectories for which all but a few of the couplings vanish. In recent years it has become increasingly apparent that renormalizability is not a fundamental physical requirement, and any realistic quantum field theory may contain non-renormalizable as well as renormalizable interactions. In fact, the experimental success of The Standard Model merely imposes a constraint on the characteristic energy scale of any non-renormalizable interaction. In particular, it remains unclear what is the fundamental principle behind the choice of the trajectory corresponding to the real world from the infinite number of possible theories.

One of the admissible ways to address this problem is to demand that the Hamiltonian of the theory belongs to a trajectory which terminates at a fixed point in UV. Theories with this property are called asymptotically safe. This concept was originally introduced by Weinberg \cite{Weinberg:1976xy,Weinberg:1979}
as an approach to select physically consistent quantum field theories (QFT). It provided an alternative
to the standard requirement of renormalizability. The original argument was motivated by an attempt to avoid hitting the Landau pole along the RG flow by imposing a demand that the flow of a QFT terminates at the UV fixed point. Asymptotically safe theories are not necessarily renormalizable in the usual sense. However, they are interacting QFTs with no unphysical singularities at high energies.


Just as the requirement of renormalizability leaves only few possible interaction terms 
which one is allowed to include in the Lagrangian, so does the requirement
of asymptotic safety imposes an infinite number of constraints, limiting the number
of physically acceptable theories. Indeed, a UV fixed point in general has only finite number of relevant deformations, therefore any asymptotically safe field theory is entirely specified by a finite number of parameters. Some of such theories are represented by a renormalizable field theory at low energies, while others might be non-renormalizable. 

Thus, for instance, a scalar field theory respecting $\phi\to-\phi$ symmetry rules out the possibility of any renormalizable interaction in five dimensions, yet it exhibits a non-trivial UV fixed point with just two relevant deformations in higher dimensions, and therefore it provides an example of asymptotically safe non-renormalizable quantum field theory \cite{Weinberg:1976xy}. An immediate application of the asymptotic safety 
is related to the problem of quantum gravity
\cite{Weinberg:1976xy,Weinberg:1979,Christensen:1978sc,Gastmans:1977ad}
 which is ongoing 
(see, \textit{e.g.}, \cite{Eichhorn:2017egq} and references therein for recent work).


Several techniques for studying UV properties of QFTs in particular and their RG flows
in general have been developed.
One of the approaches is to perform a dimensional continuation, and subsequently
apply a perturbative expansion around the desired integer-valued physical space-time
dimension.
This is done with the hope that the result
will have sufficiently accurate convergence behaviour to be applied to physically meaningful space-time
dimensions, such as $d=3,4,5$. In this spirit the Wilson-Fisher fixed point has originally
been derived in $4-\epsilon$ dimensions, in which case at the fixed point the coupling
constant is perturbative in $\epsilon$ \cite{Wilson:1971dc}.
In the context of asymptotic safety, gravity has been studied in
$2+\epsilon$ dimensions, see \cite{Weinberg:1979,Christensen:1978sc,Gastmans:1977ad}
for some of the early works.

Recently the $O(N)$ vector model of the scalar fields $\phi^a$, $a=1,\dots,N$, and $\sigma$
with the relevant cubic interactions $\phi^a\phi^a \sigma$ and $\sigma^3$ in $d=6-\epsilon$ dimensions
has been extensively studied perturbatively in $\epsilon$, see e.g., \cite{Fei:2014yja,Fei:2014xta}. This model exhibits an IR fixed point, and it was suggested as an alternative description of the critical scalar $O(N)$ model with quartic interaction $(\phi^a\phi^a)^2$.
Several consistency checks for this proposal were given \cite{Fei:2014yja,Fei:2014xta}.
Some of them, such as matching of the coefficients
of the three-point functions $\langle\phi^a\phi^b \sigma\rangle$
and $\langle\sigma\sigma\sigma\rangle$ in both theories, fall within a
universal class of results obtained for a generic large-$N$ CFT
 by the methods of conformal bootstrap \cite{Petkou:1995vu,Petkou:1994ad}.
However, some other checks,
such as matching of the anomalous dimensions of certain operators
in these theories to high order both in $\epsilon$ and $1/N$ expansion,
are rather non-trivial \cite{Fei:2014yja,Fei:2014xta}.
Recent work \cite{Giombi:2019upv} is also dedicated to calculation of non-perturbative
imaginary-valued contributions to the scaling dimensions
as a result of fluctuations around instanton background.


Large-$N$ expansion suggests another useful and widely applied tool to study non-perturbative aspects of QFTs.
A significant part of the large $N$ lore is due to
the original observation made by 't~Hooft \cite{tHooft:1973alw}
that the $SU(3)$ QCD becomes more tractable when generalized
to the $SU(N)$ gauge theory with matter, and considered in the large-$N$
limit (see also \cite{Witten:1979kh} and references therein for some of the earlier
arguments regarding the accuracy of such an approximation).
In the context of the AdS/CFT correspondence
the large-$N$ limit of the $SU(N)$ gauge theories has been subsequently related to
the weak Newton/string coupling limit of the dual bulk theory in the AdS space
\cite{Maldacena:1997re,Witten:1998qj,Gubser:1998bc}.
Other recent applications of the large-$N$ formalism include studies of the large-$N$ three-dimensional
$SU(N)$ gauge theories with the Chern-Simons interaction, and generalizations of these models to study
fundamental vector matter at finite temperature and chemical potential (see, \textit{e.g.},
 \cite{Giombi:2011kc,Aharony:2012ns,Yokoyama:2012fa,Jain:2014nza,Geracie:2015drf,Goykhman:2016zgd}
and references therein).\footnote{For recent advances in the large-$N$ QED, see \textit{e.g.}, \cite{Giombi:2016fct}.} 

Some of the earlier ideas related to applications of the
large-$N$ methods are due to the work by Parisi (see \cite{Parisi:1975} and references therein),
who in particular developed a systematic proof of renormalizability of the 
$O(N)$ vector model in $4<d<6$ dimensions to each order in the $1/N$ expansion.
In particular, one can study the $O(N)$ vector model in a physically interesting dimension $d=5$.
Therefore, this model provides a useful testing ground for the formalism of asymptotic safety \cite{Weinberg:1976xy}.

Furthermore, it was shown that the large-$N$ approach can meet the dimensional continuation method  
(see, \textit{e.g.}, \cite{Fei:2014yja} for recent developments in this direction),
thanks to the simple observation that the IR Wilson-Fisher fixed point in $d=4-\epsilon$
dimensions can be analytically continued to obtain perturbative UV fixed point in $d=4+\epsilon$ dimensions,
albeit the coupling constant at that fixed point 
is negative-valued, and therefore the theory is unstable \cite{Weinberg:1976xy}.
In fact, the work of \cite{Fei:2014yja} was partially motivated by an attempt to design a UV completion of the critical  $O(N)$ vector model in $4<d<6$ dimensions.\footnote{It should be noted that the cubic $\sigma^3$ theory considered in \cite{Fei:2014yja} is still expected to run into instabilities because of the negative mode associated with fluctuations around the instanton background. See \cite{Giombi:2019upv} for the recent detailed account of the stability issues in this model.}


In this work we continue to study the $O(N)$ vector model with quartic interaction $(\phi ^a\phi^a)^2$ in $2\leq d \leq 6$ dimensions. We scrutinize renormalization in the $d=5$ case and provide an additional evidence for the asymptotic safety of the model.
Our calculations are done to the next-to-leading
order in the $1/N$ expansion, and we successfully recover the known anomalous scaling
dimensions associated with the UV CFT  \cite{Vasiliev:1981yc,Petkou:1995vu,Petkou:1994ad,Vasiliev:1981dg}.
The calculations can be readily extended to general $d$, and therefore similar conclusions hold for the IR CFT in $2<d<4$ as well as for the UV CFT in $4<d<6$.

The primary goal of this work is to derive a CFT data associated with the three-point functions $\langle \phi^a\phi^b s\rangle$ and $\langle s s s\rangle$ to the next-to-leading order in the $1/N$ expansion.
In order to find the  $\langle sss\rangle$ three-point
function at the  ${\cal O}(1/N^{3/2})$ order we calculate the 4-loop triangle diagram, and the associated 3-loop trapezoid
diagram, as well as the three-loop bellows diagram, in general dimension $d$. To the best of our knowledge analytic expressions for these diagrams were not presented in the literature before.
The results for the $\langle \phi^a\phi^b s\rangle$ agree with
\cite{Petkou:1995vu,Petkou:1994ad}, whereas the ${\cal O}(1/N^{3/2})$ result for $\langle sss \rangle$  is new. This data is important to carve a CFT which describes the critical vector model in general dimension.  
Setting $d=6-\epsilon$ we compare our findings
with their counterparts obtained  in \cite{Fei:2014yja,Fei:2014xta} for the critical cubic model.
We find that the OPE coefficients of the $\langle sss \rangle$ 
three-point functions of these CFTs match at the next-to-leading order in the  $1/N$
expansion. We discuss the significance of this non-trivial match in the context of the non-Lagrangian
bootstrap approach to the large-$N$ CFTs with $O(N)$ symmetry.



The rest of this paper is organized as follows. In section \ref{sec:setup} we
define the model studied in this work and set our notation.
Next we assume existence of the fixed point and derive a general relation between the counterterms at the fixed point. In what follows this relation is used as a consistency check for  the existence of the fixed point.
In section \ref{sec:d5fixed point} we focus on the model in 5D. We calculate various counterterms and provide additional evidence in favor of the asymptotic safety of the model.
In section \ref{sec:propagators renormalizaion} we review diagrammatic and
computational techniques for a CFT in general $d$, and use them to evaluate various
anomalous dimensions and amplitudes of the two-point functions.
In section~\ref{sec:three-point} we calculate the three-point functions associated with a CFT that emerges at the fixed point of the model in $2\leq d \leq 6$ dimensions. We discuss our results in section \ref{sec:discussion}.

\section{Setup}
\label{sec:setup}

In this paper we focus on a $d$-dimensional Euclidean vector model governed by the bare action
\begin{equation}
\label{starting action}
S = \int  d^dx \, \left( \frac{1}{2} \, (\partial _ \mu \phi ) ^ 2 + \frac{1}{2} \, g_2   \, \phi ^ 2
+ \frac{g_4}{N} \left( \phi^ 2 \right) ^2 \right)\,,
\end{equation}
where the field $\phi$ has $N$ components, but we suppress the vector index for brevity. We keep the dimension $d$ general most of the time, however, some of the specific calculations are carried out in $d=5$ and $d=6-\epsilon$. The  behaviour of the model at the fixed point is our main objective, but for now we keep the mass parameter $g_2$ to preserve generality.

The straightforward and well-known approach (sometimes referred to as the Hubbard-Stratonovich transformation)
to study the model such as (\ref{starting action}) is to introduce the auxiliary field $s$ which has no impact on the original path integral,
\begin{equation}
\label{starting action intermediate}
S = \int  d^dx \, \left( \frac{1}{2} \, (\partial _ \mu \phi ) ^ 2 + \frac{1}{2} \, g_2   \phi ^ 2
+ \frac{g_4}{N} \left( \phi  ^ 2 \right) ^2
- \frac{1}{4g_4}\left( s - \frac{2g_4}{\sqrt{N}} \, \phi^2 \right)^2 \right)\,.
\end{equation}
After a straightforward simplification this action takes the form
\begin{equation}
\label{main action}
S = \int  d^dx \, \left( \frac{1}{2} \, (\partial _ \mu \phi ) ^ 2 + \frac{1}{2} \, g_2   \phi^ 2
- \frac{1}{4g_4} \, s ^ 2 + \frac{1}{\sqrt{N}} \, s   \phi^ 2 \right)\,.
\end{equation}
As usual, the model (\ref{main action}) has to be renormalized. 

To begin with, we introduce the renormalized fields $\tilde \phi$, $\tilde s$
\begin{equation}
\phi = \sqrt{Z_\phi} \, \tilde \phi \,, \qquad s = \sqrt{Z_s} \, \tilde s \,,
\end{equation}
where the field strength renormalization constants $Z_{\phi, s}$ are expressed in terms of the
counterterms $\delta _{\phi, s}$ as 
\begin{equation}
\label{Z phi and s definition}
Z_\phi = 1 + \delta _\phi\,, \qquad Z_s = 1 + \delta _ s\,. 
\end{equation}
Similarly, the renormalized mass $m$ and coupling $\tilde g_4$ are defined by
\begin{align}
g_2 Z_\phi &= m^2 +\delta _m\,,\\
g_4 Z_\phi ^ 2 &= \tilde g_4 +\delta _4\,,\label{definition of renormalized g4}
\end{align}
where $\delta _m$ and $\delta _4$ are mass and quartic coupling counterterms respectively.
\footnote{It follows from the definition (\ref{definition of renormalized g4}) that the interaction term in the
original bare action (\ref{starting action}) can be expressed as the following sum
\begin{equation}
g_4 ( \phi^ 2 ) ^2
=\tilde g_4 (  \tilde\phi^ 2 ) ^2
+\delta_4 (  \tilde\phi  ^2 ) ^ 2\,.
\end{equation}}

It turns out that all counterterms vanish to leading order in the $1/N$ expansion. Therefore to carry out calculations to the next-to-leading order, it is sufficient to linearize the full action with respect to various counterterms. The action (\ref{main action}) in terms of the renormalized parameters is thus given by
\begin{align}
\label{renormalized action preliminary}
S &= \int  d^dx \, \left( \frac{1}{2} \, (\partial _ \mu \tilde\phi ) ^ 2
 + \frac{1}{2} \, m^2   \tilde\phi ^ 2 
 + \frac{1}{2} \,\delta_\phi\,  (\partial _ \mu \tilde\phi ) ^ 2
 + \frac{1}{2} \,\delta_m \tilde\phi ^ 2 \right.\notag\\
&- \left.\frac{1}{4\tilde g_4} \,\left(1 + 2\delta_\phi + \delta _s - \frac{\delta _ 4}{\tilde g_4}\right) \, \tilde s ^ 2 
+ \frac{1}{\sqrt{N}} \, \left(1 + \frac{1}{2}\delta_s +\delta_\phi\right) \, \tilde s \,  \tilde \phi ^ 2 \right)\,.
\end{align}
It is convenient to define the following combinations of the counterterms,
\begin{align}
\label{definition of hat delta s}
\hat \delta _s &= 2\delta_\phi + \delta _s - \frac{\delta _ 4}{\tilde g_4}\,,\\
\label{definition of hat delta 4}
\frac{\hat \delta _4}{\sqrt{\tilde g_4}} &= \frac{1}{2}\delta_s +\delta_\phi\,,
\end{align}
in terms of which the action can be rewritten as follows
\begin{align}
\label{renormalized action}
S &= \int  d^dx \, \left( \frac{1}{2} \, (\partial _ \mu \tilde\phi ) ^ 2
 + \frac{1}{2} \, m^2  \tilde\phi  ^ 2 
 -\frac{1}{4\tilde g_4} \, \tilde s ^ 2 
+ \frac{1}{\sqrt{N}} \,  \tilde s \,  \tilde \phi  ^ 2
 \right.\notag\\
&+ \left.  \frac{1}{2} \,\delta_\phi\,  (\partial _ \mu \tilde\phi ) ^ 2
 + \frac{1}{2} \,\delta_m \tilde\phi ^ 2 
-\frac{1}{4\tilde g_4} \, \hat\delta _s \, \tilde s ^ 2 
+ \frac{1}{\sqrt{N}} \, \frac{\hat \delta _4}{\sqrt{\tilde g_4}} \, \tilde s \, \tilde \phi ^ 2 \right)\,.
\end{align}
Loop corrections along with renormalization conditions fix the counterterms $\delta_\phi$, $\delta _m$, $\hat \delta _s$, and $\hat \delta _4$. Inverting (\ref{definition of hat delta s}), (\ref{definition of hat delta 4}) one then obtains
\begin{align}
\label{delta s in terms of auxiliaries}
\delta _ s & = 2\left(\frac{\hat\delta _ 4 }{\sqrt{ \tilde g _ 4 }} - \delta _ \phi\right)\,, \\
\label{delta 4 in terms of auxiliaries}
\frac{\delta _4}{\tilde g_4} &= 2\frac{\hat \delta _4}{\sqrt{\tilde g_4}} - \hat \delta _s\,.
\end{align}
Finally, renormalization of the quartic coupling constant can be read off (\ref{Z phi and s definition}),  (\ref{definition of renormalized g4}) and (\ref{delta 4 in terms of auxiliaries}) 
\begin{equation}
\label{g4 renormalization}
g_4=\tilde g_4 \left( 1+2\frac{\hat\delta _ 4}{\sqrt{\tilde g_4}} - 2 \delta _ \phi -\hat \delta _s \right)
=\tilde g_4 \left( 1+\delta_s-\hat \delta _s \right)\,,
\end{equation}
where in the last equality  (\ref{delta s in terms of auxiliaries}) was used. 

The Callan-Symanzik equation can be used to find various relations between the counterterms of the theory. In particular, using it for the three-point function $\langle \phi\phi s\rangle$, we argue in Appendix \ref{appendix:callan-symanzik equation} that at the UV (IR) fixed point in $4<d<6$ ($2<d<4$) dimensions one obtains
\begin{equation}
\label{delta s hat delta s result}
\mu\frac{\partial}{\partial\mu} \left(\delta _s - \hat \delta_s +2 B {\cal L}_s \right)=0+\mathcal{O}(1/N^2) \,,
\end{equation}
where $\mu$ is an arbitrary renormalization scale.
In section \ref{sec:d5fixed point} we explicitly confirm this relation in $d=5$ at the next-to-leading order in the $1/N$ expansion. We interpret it as a consistency check of the assumption that the theory is asymptotically safe. 

Of course, the Callan-Symanzik equation is just a statement about consistency of the calculations at various scales of RG flow, and therefore it holds in any field theory. However, it depends on the beta functions of the couplings which are not known in general dimension. For instance, existence of the fixed point in the $\phi^4$ vector model was not verified directly in general $d$. This is just an assumption based on accumulated evidence rather than a derived fact.  In particular, eq. (\ref{delta s hat delta s result}) is the Callan-Symanzik equation with beta functions suppressed by hand. Yet, it is satisfied by the remaining counterterms which are calculated separately.  In this sense (\ref{delta s hat delta s result}) plays a role of the consistency check for the existence of a fixed point.

\section{UV fixed point in $d=5$}
\label{sec:d5fixed point}

A vast amount of literature, starting from the earlier works \cite{Parisi:1975,Weinberg:1976xy},
is dedicated to 
studying the properties of the UV conformal fixed point of the $O(N)$ vector models with quartic interaction
in $4<d<6$.
While the quartic interaction is non-renormalizable in $d>4$, one can take advantage
of the fact that this model is renormalizable at each order in the $1/N$ expansion \cite{Parisi:1975}.
In this section we consider (\ref{renormalized action}) 
in $d=5$ dimensions. The main goal is to systematically derive various anomalous dimensions
\cite{Vasiliev:1981yc,Vasiliev:1981dg,Petkou:1995vu,Petkou:1994ad} and to verify explicitly
that they satisfy certain constraints which are expected to hold at the fixed point.
The calculations are carried out to the next-to-leading order in the $1/N$ expansion.

To begin with, we list all necessary Feynman rules for the model  (\ref{renormalized action}).
A solid line will be used to denote the propagators of various components of the scalar field $\phi$, whereas a
dashed line is associated with the propagator of the Hubbard-Stratonovich auxiliary field $s$. Obviously, the matrix of propagators is diagonal, whereas each non-trivial interaction vertex carries two identical vector indices. Hence, for simplicity we suppress the Kronecker delta which explicitly emphasizes these facts. We also omit the momentum conservation delta function at each vertex. Vertices are denoted by a solid blob,
whereas propagators are enclosed by black dots (absence of dots indicates an amputated external leg).
\begin{center}
  \begin{picture}(700,155) (0,0)
    \SetWidth{1.0}
    \SetColor{Black}
    \Vertex(30,150){2}
    \Text(55,155)[lb]{$p$}
    \Line[](30,150)(80,150)
    \Vertex(80,150){2}
    \Text(95,142)[lb]{$=\frac{1}{p^2 + m^2}$}
    \Vertex(250,150){2}
    \Text(275,155)[lb]{$p$}
    \Line[dash,dashsize=10](250,150)(300,150)
    \Vertex(300,150){2}
    \Text(315,145)[lb]{$=-2\tilde g_4$}
    \Text(55,115)[lb]{$p$}
    \Line[](30,105)(50,105)
    \Arc[](55,105)(5,135,495)
    \Line[](52,101)(58,109)
    \Line[](52,109)(58,101)
    \Line[](60,105)(80,105)
    \Text(95,100)[lb]{$=-\delta_\phi p^2-\delta _m$}
    \Line[dash,dashsize=10](250,105)(290,105)
    \Arc[](295,105)(5,135,495)
    \Line[](292,101)(298,109)
    \Line[](292,109)(298,101)
    \Text(315,100)[lb]{$=\delta _0$}
    \Line[dash,dashsize=10](30,40)(80,40)
    \Vertex(80,40){4}
    \Line[](80,40)(110,70)
    \Line[](80,40)(110,10)
     \Text(125,30)[lb]{$=-\frac{2}{\sqrt{N}}$}
    \Line[dash,dashsize=10](250,40)(290,40)
    \Arc[](295,40)(5,135,495)
    \Line[](292,36)(298,44)
    \Line[](292,44)(298,36)
    \Line[](298,44)(330,70)
    \Line[](298,36)(330,10)
     \Text(345,30)[lb]{$=-\frac{2}{\sqrt{N}}\frac{\hat \delta _4}{\sqrt{\tilde g_4}}$}
  \end{picture}
\end{center}
Note that we introduced a counterterm $\delta _0$. It is completely determined by the renormalization condition $\langle \tilde s \rangle=0$. In what follows we assume this condition has been satisfied without elaborating the details. 
 
The full propagator of the Hubbard-Stratonovich field $\tilde s$
to leading order in the $1/N$ expansion is thus given by an infinite sum of bubble diagrams
\begin{center}
  \begin{picture}(670,40) (105,-28)
    \SetWidth{1.0}
    \SetColor{Black}
    \Vertex(120,-4){2}
    \Line[dash,dashsize=10](120,-4)(170,-4)
    \Vertex(170,-4){2}
    \Text(186,-8)[lb]{$+$}
    \Vertex(210,-4){2}
    \Line[dash,dashsize=10](210,-4)(230,-4)
    \Vertex(230,-4){4}
    \Arc[](255,-4)(25,135,495)
    \Vertex(280,-4){4}
    \Line[dash,dashsize=10](280,-4)(300,-4)
    \Vertex(300,-4){2}
    \Text(316,-8)[lb]{$+$}
    \Vertex(340,-4){2}
    \Line[dash,dashsize=10](340,-4)(360,-4)
    \Vertex(360,-4){4}
    \Arc[](385,-4)(25,135,495)
    \Vertex(410,-4){4}
    \Line[dash,dashsize=10](410,-4)(430,-4)
    \Vertex(430,-4){4}
    \Arc[](455,-4)(25,135,495)
    \Vertex(480,-4){4}
    \Line[dash,dashsize=10](480,-4)(500,-4)
    \Vertex(500,-4){2}
    \Text(516,-8)[lb]{$+\dots $}
  \end{picture}
\end{center} 
This is a geometric series which can be readily written in a closed form.\footnote{One should take into account that each bubble comes with the 
symmetry factor $1/2$.} We denote it by $G_s(p)$ and represent it diagrammatically by a wavy line 
\begin{center}
  \begin{picture}(670,10) (10,-10)
    \SetWidth{1.0}
    \SetColor{Black}
    \Vertex(30,-4){2}
    \Photon(30,-5)(80,-4){2}{9}
    \Text(55,5)[lb]{$p$}
    \Vertex(80,-4){2}
    \Text(95,-12)[lb]{$= G_s(p)=-2\tilde g_4\sum_{n=0}^\infty (-4\tilde g_4 B) ^ n=\frac{-2\tilde g_4}
{1+4\tilde g_4 B}~,$}
  \end{picture}
\end{center} 
where each bubble in the infinite series is associated with a UV divergent loop integral
\begin{equation}
\label{B definition}
B(p) = \int \frac{d^5q}{(2\pi)^5}\,\frac{1}{(q^2+m^2)((p+q)^2 + m^2)}\,.
\end{equation} 
To regularize the UV divergence we introduce a spherically symmetric sharp cutoff $\Lambda$
\begin{equation}
\label{B0 result}
B(p) = \frac{\Lambda}{12\pi^3} -\frac{1}{64\pi^2 p}\,\left((4m^2+p^2)\,\tan^{-1}\left(\frac{p}{2|m|}\right)+2|m|p\right)\,.
\end{equation} 
The power law divergence can be eliminated by adjusting the counterterm $\hat\delta _s$. Unlike logarithmic divergences, the power law divergences depend on the details of regularization scheme. For instance, they are absent in dimensional regularization. Hence, we simply ignore such divergences in what follows to reduce clutter in the equations. In particular, in the large momentum limit which we are interested in for the purpose of finding the UV fixed point, we obtain
\begin{equation}
\label{B0 in UV}
B\Big|_{p\gg m} = -\frac{p}{128\pi}\,.
\end{equation}
Using the asymptotic behaviour (\ref{B0 in UV}) rather than the full expression (\ref{B0 result}), and thus focusing on the UV
regime, one avoids passing through the pole of the propagator $G_s(p)$ at some finite momentum 
\cite{Parisi:1975}. Therefore in all of our calculations
we take the limit $\tilde g_4 p \gg 1$ to simplify the propagator of the Hubbard-Stratonovich field
(see, \textit{e.g.}, \cite{Fei:2014yja}), 
\begin{equation}
\label{massless full s propagator}
G_s(p)=-\frac{1}{2B(p)} = \frac{64\pi}{p}\,.
\end{equation}

By assumption, the beta function for $\tilde g_4$ is expected to exhibit a UV fixed point. As pointed out in Appendix  \ref{appendix:callan-symanzik equation}, the counterterms must therefore satisfy equation (\ref{delta s hat delta s result}). In particular, (\ref{delta s hat delta s result}) serves as a non-trivial consistency check of the assumption about the UV behaviour of the beta function. In order to explicitly verify this identity, we proceed to calculation of the counterterms to ${\cal O}\left(\frac{1}{N}\right)$ order. It is natural to set $m^2=0$ since we are ultimately interested in the UV behaviour of the loop integrals.

\subsection{Scalar field two-point function}

To leading order in the $1/N$ expansion only $G_s(p)$ exhibits loop corrections. As argued above, there are no non-trivial UV divergences associated with loop diagrams at this order, and all counterterms thus vanish. However,  a
non-trivial renormalization is induced at the next-to-leading order in $1/N$.

The ${\cal O}\left(\frac{1}{N}\right)$ contribution to the counterterms $\delta_{\phi,m}$ are derived from the requirement
that the sum of the loop diagram 
\begin{center}
  \begin{picture}(670,60) (125,-40)
    \Vertex(180,-4){2}
    \Line[](180,-4)(230,-4)
    \Text(205,0)[lb]{$p$}
    \Vertex(230,-4){4}
    \Arc[](255,-4)(25,-180,-360)
    \Text(245,-45)[lb]{$p+q$}
    \PhotonArc[](255,-4)(25,0,180){2}{9}
    \Text(255,30)[lb]{$q$}
    \Vertex(280,-4){4}
    \Line[](280,-4)(330,-4)
    \Text(305,0)[lb]{$p$}
    \Vertex(330,-4){2}
    \Text(340,-15)[lb]{$=\frac{1}{(p^2)^2}\left(-\frac{2}{\sqrt{N}}\right)^2\int \frac{d^5q}{(2\pi)^5}\frac{-1}{2B(q)}\,\frac{1}{(p+q)^2}$}
  \end{picture}
\end{center}
and the counter-term contribution 
\begin{center}
  \begin{picture}(670,40) (315,-30)
    \Vertex(370,-4){2}
    \Line[](370,-4)(420,-4)
    \Arc[](425,-4)(5,135,495)
    \Line[](422,-8)(428,0)
    \Line[](422,0)(428,-8)
    \Line[](430,-4)(480,-4)
    \Vertex(480,-4){2}
    \Text(395,2)[lb]{$p$}
    \Text(455,2)[lb]{$p$}
    \Text(495,-13)[lb]{$=\frac{1}{(p^2)^2}\,(-p^2\delta_\phi -\delta _m)$}
  \end{picture}
\end{center}
is finite. 
Introducing a sharp cutoff $\Lambda$ and focusing on the logarithmic divergences only, yields\footnote{If we keep the mass in the action (\ref{renormalized action}), then there is an additional counterterm of the form 
\begin{equation} 
\delta _ m = -\frac{1}{N}\frac{320m^2}
{3\pi^2}\,\log\left(\frac{\Lambda}{\mu}\right)\,. 
\end{equation} 
}
\begin{align}
\label{delta phi d5}
\delta _ \phi &= -\frac{1}{N}\frac{64}{15\pi^2}\,\log\left(\frac{\Lambda}{\mu}\right)\,,
\end{align}
where $\mu$ is an arbitrary renormalization scale. The anomalous dimension of the scalar field $\phi$ at the UV fixed point can be readily evaluated using the Callan-Symanzik equation for the two-point function
of $\tilde\phi$, see  Appendix  \ref{appendix:callan-symanzik equation}
\begin{equation}
\label{gamma_phi d=5 result}
\gamma_\phi = \frac{1}{2}\frac{\partial}{\partial\log\mu}\,\delta_\phi = 
\frac{1}{N}\frac{32}{15\pi^2}\,.
\end{equation}
This result is in full agreement with \cite{Vasiliev:1981yc,Vasiliev:1981dg,Petkou:1995vu,Petkou:1994ad}.

\subsection{Renormalization of the interaction vertex}
\label{sec:interaction vertex in d=5}

There are two loop diagrams that contribute to renormalization of the interaction vertex at the next to leading order in the $1/N$ expansion. To calculate the corresponding counterterm $\hat \delta _4$ it is enough to set all external momenta to zero. The first diagram takes the form
\begin{center}
  \begin{picture}(450,194) (55,-63)
    \SetWidth{1.0}
    \SetColor{Black}
    \Line[](96,114)(176,34)
    \Line[](176,34)(96,-46)
    \Photon(128,82)(128,-14){2}{9}
    \Text(115,30)[lb]{$q$}
    \Text(155,0)[lb]{$q$}
    \Text(155,60)[lb]{$q$}
    \Vertex(128,82){4}
    \Vertex(128,-14){4}
    \Line[dash,dashsize=10](176,34)(208,34)
    \Vertex(176,34){4}
      \Text(226, 23)[lb]{$ =\left(-\frac{2}{\sqrt{N}}\right)^3 \int \frac{d^5q}{(2\pi)^5}\frac{-1}{2B(q)}\,\frac{1}{(q^2)^2}
                                    =-\frac{1}{N^{3/2}}\,\frac{128}{3\pi^2}\,\log\Lambda$}
  \end{picture}
\end{center}
whereas the second diagram is given by
\begin{center}
  \begin{picture}(400,194) (329,-63)
    \SetWidth{1.0}
    \SetColor{Black}
    \Line[](384,66)(384,2)
    \Text(356,30)[lb]{$p+q$}
    \Text(340,30)[lb]{$q$}
    \Text(373,85)[lb]{$q$}
    \Text(373,-25)[lb]{$q$}
    \Text(373,85)[lb]{$q$}
    \Text(405,53)[lb]{$p$}
    \Text(405,8)[lb]{$p$}
    \Line[](384,2)(416,34)
    \Line[](384,66)(416,34)
    \Vertex(416,34){4}
    \Line[dash,dashsize=10](416,34)(448,34)
    \Vertex(352,98){4}
    \Photon(352,98)(384,66){2}{9}
    \Vertex(384,66){4}
    \Vertex(384,2){4}
    \Photon(384,2)(352,-30){2}{9}
    \Vertex(352,-30){4}
    \Line[](352,98)(320,130)
    \Line[](352,-30)(320,-62)
    \Line[](352,98)(352,-30)
      \Text(460, 23)[lb]{$= 
        \left(-\frac{2}{\sqrt{N}}\right)^5 N \int \frac{d^5q}{(2\pi)^5}\left(\frac{-1}{2B(q)}\right)^2\,                      
         \frac{1}{q^2}\,\left(-\frac{1}{2}\right)
         \frac{\partial B(q)}{\partial m^2} =-\frac{1}{N^{3/2}}\,\frac{512}{3\pi^2}\,\log\Lambda
         ~,$}
  \end{picture}
\end{center}
where we used (\ref{B definition}) to get the following simple relation 
\begin{equation}
-\frac{1}{2}\,\frac{\partial B(q)}{\partial m^2}= \int \frac{d^5p}{(2\pi)^5}\,\frac{1}{(p^2+m^2)^2((p+q)^2 + m^2)}\,.
\end{equation}
In particular, the loop integral can be calculated by taking derivative of (\ref{B0 result}) with respect to the mass parameter.

Finally, the tree level counterterm contribution can be read off the Feynman rules we listed in the beginning of this section. 
Combining everything together and demanding cancellation of the divergences, we obtain
\begin{equation}
\label{hat delta 4 result}
\frac{\hat \delta _4}{\sqrt{\tilde g_4}} = -\frac{1}{N}\frac{320}{3\pi^2}\,
\,\log\left(\frac{\Lambda}{\mu}\right)\,.
\end{equation}

\subsection{Auxiliary field propagator}

Next let us evaluate the counterterm $\hat\delta _s$. To this end, we perform renormalization of the Hubbard-Stratonovich propagator. There are five loop diagrams which contribute at the next to leading order in the $1/N$ expansion. We tag these diagrams with $C_{i}$, $i=1,\dots,5$ and denote their external momentum by $u$. For simplicity, all the diagrams are amputated, \textit{i.e.}, the two external $s$ propagators are factored out. Thus, for instance
\begin{center}
  \begin{picture}(194,80) (31,-25)
    \SetWidth{1.0}
    \SetColor{Black}
    \Text(-5,5)[lb]{$C_1=$}
    \Line[dash,dashsize=10](32,10)(94,10)
    \Text(60,15)[lb]{$u$}
    \Arc[](128,10)(35,153,513)
    \Line[dash,dashsize=10](162,10)(224,10)
    \Text(190,15)[lb]{$u$}
    \Photon(128,44)(128,-24){2}{9}
    \Text(117,10)[lb]{$r$}
    \Text(70,35)[lb]{$p+u$}
    \Text(90,-20)[lb]{$p$}
    \Text(160,-20)[lb]{$p+r$}
    \Text(160,35)[lb]{$p+r+u$}
    \Vertex(94,10){4}
    \Vertex(162,10){4}
    \Vertex(128,-24){4}
    \Vertex(128,44){4}
  \end{picture}
\end{center}
is given by
\begin{equation}
C_1 =\frac{1}{2}\left(-\frac{2}{\sqrt{N}}\right)^4 N
\int\frac{d^5r}{(2\pi)^5}\,\frac{64\pi}{r}
\int\frac{d^5p}{(2\pi)^5}
\frac{1}{p^2(p+u)^2(p+r)^2(p+r+u)^2}\,.
\end{equation}
This integral is somewhat laborious to evaluate in full generality. However, we are only interested in its behaviour at large momenta
\begin{equation}
\label{C1 result}
C_1 = \frac{1024\pi}{N}
\int\frac{d^5p}{(2\pi)^5}
\frac{1}{p^2(p+u)^2}
\int\frac{d^5r}{(2\pi)^5}\,\frac{1}{r^5}
=\frac{1}{N}\frac{256}{3\pi^2}\,B(u)\,\log\,\Lambda+\dots\,.
\end{equation}
Similarly,
\begin{center}
  \begin{picture}(322,100) (31,-33)
    \SetWidth{1.0}
    \SetColor{Black}
    \Text(-5,11)[lb]{$C_2=$}
    \Text(60,22)[lb]{$u$}
    \Text(323,22)[lb]{$u$}
    \Text(100,-22)[lb]{$p$}
    \Text(275,-22)[lb]{$r$}
    \Text(280,45)[lb]{$r+u$}
    \Text(80,45)[lb]{$p+u$}
    \Text(190,55)[lb]{$q$}
    \Text(180,-30)[lb]{$q+u$}
    \Text(113,11)[lb]{$p+q+u$}
    \Text(225,11)[lb]{$r+q+u$}
    \Line[dash,dashsize=10](32,16)(94,16)
    \Arc[](128,16)(35,153,513)
    \Photon(144,48)(240,48){2}{9}
    \Photon(144,-16)(240,-16){2}{9}
    \Arc[](256,16)(35,153,513)
    \Line[dash,dashsize=10](292,16)(354,16)
    \Vertex(94,16){4}
    \Vertex(292,16){4}
    \Vertex(144,48){4}
    \Vertex(240,48){4}
    \Vertex(144,-16){4}
    \Vertex(240,-16){4}
  \end{picture}
\end{center}
is given by
\begin{equation}
C_2 =\frac{1}{2}\left(-\frac{2}{\sqrt{N}}\right)^6 N^2
\int\frac{d^5q}{(2\pi)^5}\,\frac{64\pi}{q}
\,\frac{64\pi}{|q+u|}
\left(\int\frac{d^5p}{(2\pi)^5}\frac{1}{p^2(p+u)^2(p+q+u)^2}\right)^2\,.
\end{equation}
Expanding it in the region of large momenta, gives
\begin{align}
C_2&=\frac{2^{18}\pi^2}{N}
\int\frac{d^5p}{(2\pi)^5}
\frac{1}{p^2(p+u)^2}
\int\frac{d^5r}{(2\pi)^5}\frac{1}{(r^2)^2}
\int\frac{d^5q}{(2\pi)^5}\frac{1}{(q^2)^2(r+q)^2}\notag\\
&=\frac{1}{N}\frac{1024}{3\pi^2}\,B(u)\,\log\,\Lambda+\dots\,.
\label{C2 result}
\end{align}
Next, we focus on
\begin{center}
  \begin{picture}(194,100) (31,-35)
    \SetWidth{1.0}
    \SetColor{Black}
    \Text(-5,5)[lb]{$C_3=$}
    \Line[dash,dashsize=10](32,10)(94,10)
    \Text(60,15)[lb]{$u$}
    \Text(85,15)[lb]{$p$}
    \Text(168,15)[lb]{$p$}
    \Arc[](128,10)(35,153,513)
    \Line[dash,dashsize=10](162,10)(224,10)
    \Text(190,15)[lb]{$u$}
    \Photon(97,25)(159,25){2}{9}
    \Text(125,13)[lb]{$q$}
    \Text(115,50)[lb]{$p+q$}
    \Text(115,-38)[lb]{$p+u$}
    \Vertex(94,10){4}
    \Vertex(162,10){4}
    \Vertex(97,25){4}
    \Vertex(159,25){4}
  \end{picture}
\end{center}
Using Feynman rules results in the following integral expression
\begin{equation}
C_3 =\left(-\frac{2}{\sqrt{N}}\right)^4 N
\int\frac{d^5p}{(2\pi)^5}\,\frac{1}{(p^2)^2(p+u)^2}
\int\frac{d^5q}{(2\pi)^5}\,\frac{64\pi}{q(p+q)^2}\,.
\end{equation}
Performing integration over $q$ and keeping the logarithmic divergence only\footnote{Recall that we ignore scheme dependent power law divergences.}, yields
\begin{equation}
\label{C3 result}
C_3=-\frac{1}{N}\frac{256}{15\pi^2}\,B(u)\,\log\,\Lambda+\dots\,.
\end{equation}
The two remaining diagrams are
\begin{center}
  \begin{picture}(194,90) (31,-27)
    \SetWidth{1.0}
    \SetColor{Black}
    \Text(-15,5)[lb]{$C_4=2\times$}
    \Line[dash,dashsize=10](32,10)(89,10)
    \Text(60,15)[lb]{$u$}
    \Arc[](128,10)(35,0,173)
    \Arc[](128,10)(35,187,360)
    \Line[dash,dashsize=10](162,10)(224,10)
    \Text(190,15)[lb]{$u$}
    \Text(125,50)[lb]{$p$}
    \Text(115,-38)[lb]{$p+u$}
    \Arc[](94,10)(5,135,495)
    \Line[](91,6)(97,14)
    \Line[](91,14)(97,6)
    \Vertex(162,10){4}
  \end{picture}
\end{center}
and
\begin{center}
  \begin{picture}(194,90) (31,-35)
    \SetWidth{1.0}
    \SetColor{Black}
    \Text(-5,5)[lb]{$C_5=$}
    \Line[dash,dashsize=10](32,10)(94,10)
    \Text(60,15)[lb]{$u$}
    \Arc[](128,10)(35,0,83)
    \Arc[](128,10)(35,97,360)
    \Line[dash,dashsize=10](162,10)(224,10)
    \Text(190,15)[lb]{$u$}
    \Text(125,55)[lb]{$p$}
    \Text(115,-38)[lb]{$p+u$}
    \Arc[](128,45)(5,135,495)
    \Line[](125,41)(131,49)
    \Line[](125,49)(131,41)
    \Vertex(94,10){4}
    \Vertex(162,10){4}
  \end{picture}
\end{center}
An extra factor of two in $C_4$ comes from the interchange of the ordinary vertex with the counterterm. More specifically, the integral expressions for these diagrams read
\begin{align}
\label{C4 result}
C_4 &=2\times\frac{1}{2}\left(-\frac{2}{\sqrt{N}}\right)
\left(-\frac{2}{\sqrt{N}}\frac{\hat\delta _4}{\sqrt{\tilde g_4}}\right)N
\int\frac{d^5p}{(2\pi)^5}\,\frac{1}{p^2(p+u)^2}=4\frac{\hat\delta _4}{\sqrt{\tilde g_4}}B(u)\,, \\
\label{C5 result}
C_5 &=\left(-\frac{2}{\sqrt{N}}\right)^2N
\int\frac{d^5p}{(2\pi)^5}\,\frac{-p^2\delta _\phi }{(p^2)^2(p+u)^2}=-4\delta _\phi B(u)\,.
\end{align}

Combining the loop diagrams
(\ref{C1 result}), (\ref{C2 result}), (\ref{C3 result}), (\ref{C4 result}), (\ref{C5 result})
results in the total loop correction to the $\langle \tilde s\tilde s\rangle$ propagator\footnote{$\mu_0$ is an arbitrary  IR scale.}
\begin{align}
{\cal L}_s=\left(-\frac{1}{2B}\right)^2\sum_{i=1}^5 C_i
=\frac{1}{B}\frac{1}{N}\,\frac{512}{5\pi^2}\,\log\,\frac{\mu}{\mu_{0}}\,.
\end{align}
It satisfies
\begin{equation}
\label{Ls in d=5}
B\, \mu\frac{\partial}{\partial\mu}\,{\cal L}_s =- \mu\frac{\partial}{\partial\mu}\,
\left(\frac{\hat\delta _4}{\sqrt{\tilde g_4}}-\delta_\phi\right)
\end{equation}
Notice that ${\cal L}_s$ is finite (the logarithmically divergent terms cancelled out),
and therefore
\begin{equation}
\label{hat delta s result in d=5}
\hat\delta _s=0\,.
\end{equation}
Using the relation (\ref{delta s in terms of auxiliaries}) we
can re-write (\ref{Ls in d=5}) as
\begin{equation}
\label{delta s Ls in d=5}
\mu\frac{\partial}{\partial\mu} \left(\delta _s+2 B {\cal L}_s \right)=0\,.
\end{equation}
Our calculations (\ref{hat delta s result in d=5}), (\ref{delta s Ls in d=5}) in $d=5$
agree with the general expression (\ref{delta s hat delta s result}). As discussed at the end of section \ref{sec:setup} we interpret it as a consistency check of the assumption that the model is asymptotically safe.


Finally, using the Callan-Symanzik equation for the two-point function of $\tilde s$,
see Appendix \ref{appendix:callan-symanzik equation},
we derive the anomalous dimension of the auxiliary field $\tilde s$
to ${\cal O}(1/N)$ order,
\begin{equation}
\label{gamma_s d=5 result}
\gamma_s = \frac{1}{2}\mu\frac{\partial}{\partial\mu}\,(\hat \delta_s +
2B{\cal L}_s) = 
\frac{1}{N}\frac{512}{5\pi^2}\,.
\end{equation}
This result agrees with the literature
 \cite{Vasiliev:1981yc,Vasiliev:1981dg,Petkou:1995vu,Petkou:1994ad} and
  serves as a check of the calculation.

\section{Position space calculation in general dimension}
\label{sec:propagators renormalizaion}

In the previous section we focused on providing an evidence for the existence of UV fixed point in the  
5D vector model. Our next goal is to derive a new CFT data associated with the model at criticality in $2\leq d \leq 6$. The stage is set in this section, whereas the new results are presented and derived in section \ref{sec:three-point}.

We start from calculating the next-to-leading ${\cal O}\left(\frac{1}{N}\right)$ order
corrections to the CFT propagators of the
fundamental scalar field $\phi$
and the auxiliary Hubbard-Stratonovich field $s$, working in position space at the UV fixed point. Space-time
dimension $d$ will be assumed to be general $4<d<6$ in this section, albeit
the results are also applicable to the IR conformal fixed point in $2<d<4$ dimensions.
To leading order in the large-$N$ expansion the propagators are given by
(to avoid clutter we continue to suppress the $O(N)$ vector indices and the explicit factors of the Kronecker delta)
\begin{align}
\label{free phi propagator in position space}
\langle \phi (x_1)\phi (x_2)\rangle &=
C_\phi \,\frac{1}
{|x_{12}|^{2\Delta_\phi}}\,,\\
\label{free s propagator in position space}
\langle s(x_1)s(x_2)\rangle &=C_s \,\frac{1}{|x_{12}|^{2\Delta_s}}\,,
\end{align}  
where  $x_{12}=x_1-x_2$ and the amplitudes $C_{\phi,s}$ are specified below.
To calculate the sub-leading corrections we employ the technique developed in \cite{Vasiliev:1981yc,Vasiliev:1981dg,Vasiliev:1975mq}, following recent developments in \cite{Gubser:2017vgc} (see also
 \cite{Vasiliev:1982dc,Vasilev:2004yr,Gracey:2018ame,Kotikov:2018wxe}
 for an extensive review and discussion of the $1/N$ calculations). The $1/N$ corrections dress the above propagators. As a result, their form is modified as follows
\begin{align}
\label{corrected phi propagator in position space}
\langle \phi(x_1)\phi(x_2)\rangle &=
C_\phi (1+A_\phi)\,\frac{\mu^{-2\gamma_\phi}}
{|x_{12}|^{2(\Delta_\phi + \gamma_\phi)}}\,,\\
\label{corrected s propagator in position space}
\langle s(x_1)s(x_2)\rangle &=C_s (1+A_s)\,\frac{\mu^{-2\gamma_s}}{|x_{12}|^{2(\Delta_s + \gamma_s)}}\,,
\end{align}  
where $\mu$ is an arbitrary renormalization scale, $\gamma_{\phi,s}$ represent anomalous dimensions, whereas 
the symbols $\Delta_{\phi,s}$ stand for the scaling dimensions of the fields $\phi$ and $s$ at the gaussian fixed point, 
\begin{equation}
\label{engineering scaling}
\Delta_\phi = \frac{d}{2}-1\,,\qquad \Delta_s =2 \,, \qquad 2\Delta_\phi +\Delta_s = d\, .
\end{equation}
While we have already derived the anomalous dimensions $\gamma_{\phi,s}$ 
in section \ref{sec:setup} for the five dimensional model,
in this section we calculate both the anomalous dimensions and the amplitudes $A_{\phi,s}$ in general dimension $d$. Our findings in this section
are in full agreement with the known results in the literature, \textit{e.g.}, \cite{Derkachov:1997ch},
and therefore the reader familiar with the subject can proceed to the next section. 

\subsection{Preliminary remarks}

Performing the Fourier transform
\begin{equation}
\int\frac{d^dk}{(2\pi)^d}\,e^{ik\cdot x}\frac{1}{(k^2)^{\frac{d}{2}-\Delta}}=
\frac{2^{2\Delta-d}}{\pi^{\frac{d}{2}}}\frac{\Gamma(\Delta)}{\Gamma\left(\frac{d}{2}-\Delta\right)}
\frac{1}{|x|^{2\Delta}}\,,
\end{equation}
we find the coefficient $C_\phi$ in (\ref{corrected phi propagator in position space})
\begin{equation}
C_\phi = \frac{1}{4\pi^{\frac{d}{2}}}\,\Gamma\left(\frac{d}{2}-
1\right)\,.
\end{equation}
Using expression (\ref{massless full s propagator}) for the $s$-propagator in momentum space, 
as well as the result for the bubble loop integral $B$ in general $d$ 
\begin{equation}
\label{B in general d}
B(p) = \int\frac{d^dq}{(2\pi)^d}\frac{1}{q^2(p+q)^2}
=-\frac{(p^2)^{\frac{d}{2}-2}}{2^d(4\pi)^\frac{d-3}{2}\Gamma\left(\frac{d-1}{2}\right)
\sin\left(\frac{\pi d}{2}\right)}
\end{equation}
we obtain
\footnote{It agrees with \cite{Fei:2014yja} after reconciling conventions for normalization of
the Hubbard-Stratonovich field.}
\begin{equation}
C_s=\frac{2^d\,\Gamma\left(\frac{d-1}{2}\right)\sin\left(\frac{\pi d}{2}\right)}{\pi^\frac{3}{2}
\Gamma\left(\frac{d}{2}-2\right)}\,.
\end{equation}

Let us now review the Feynman rules for a CFT in position space \cite{Vasiliev:1981yc,Vasiliev:1981dg,Vasiliev:1975mq}.
 At the conformal fixed point, there is no need to keep separate notation (solid or dashed lines)
for the propagators of $\phi$ and $s$. Instead, we put the power index $2a$
(determined by the scaling dimension $a$ of the corresponding field) on top of the line.  The line itself is assumed to be normalized to unity.
\begin{center}

  \begin{picture}(98,0) (31,-65)
    \SetWidth{1.0}
    \SetColor{Black}
    \Line[](30,-58)(90,-58)
    \Vertex(30,-58){2}
    \Vertex(90,-58){2}
    \Text(15,-58)[lb]{$x_1$}
    \Text(95,-58)[lb]{$x_2$}
    \Text(55,-53)[lb]{$2a$}
     \Text(120,-67)[lb]{$=\frac{1}{|x_{12}|^{2a}}$}
  \end{picture}

\end{center}
In particular, the calculation of each diagram starts from counting and writing down explicitly all the factors of
$C_{\phi,s}$. The loop diagram in position space simply amounts to adding two powers together.
\begin{center}

  \begin{picture}(257,50) (0,0)
    \SetWidth{1.0}
    \SetColor{Black}
    \Arc[clock](81,-39)(77.006,127.614,52.386)
    \Arc[](80,86)(80,-126.87,-53.13)
    \Line[](160,22)(256,22)
    \Vertex(160,22){2}
    \Vertex(256,22){2}
    \Vertex(33,22){2}
    \Vertex(129,22){2}
    \Text(142,20)[lb]{$=$}
    \Text(77,-5)[lb]{$2b$}
    \Text(77,43)[lb]{$2a$}
    \Text(190,27)[lb]{$2(a+b)$}
  \end{picture}

\end{center}
We also need the propagator splitting/merging relation \cite{Vasiliev:1981yc,Vasiliev:1981dg,Vasiliev:1975mq}
\begin{equation}
\label{propagator splitting}
\int d^d x_3\, \frac{1}{(x_3^2)^a((x_3-x_{12})^2)^b}
=U(a,b,d-a-b)\,\frac{1}{(x_{12}^2)^{a+b-\frac{d}{2}}}\,,
\end{equation}
where 
\begin{equation}
\label{U definition}
U(a,b,c) = \pi^\frac{d}{2}
\,\frac{\Gamma\left(\frac{d}{2}-a\right)\Gamma\left(\frac{d}{2}-b\right)\Gamma\left(\frac{d}{2}-c\right)}
{\Gamma(a)\Gamma(b)\Gamma(c)}\,.
\end{equation}
Expression (\ref{propagator splitting}) can be graphically represented as 
\begin{center}

  \begin{picture}(98,50) (130,-80)
    \SetWidth{1.0}
    \SetColor{Black}
    \Line[](30,-58)(90,-58)
    \Line[](90,-58)(150,-58)
    \Line[](180,-58)(240,-58)
    \Vertex(30,-58){2}
    \Vertex(90,-58){4}
    \Vertex(150,-58){2}
    \Vertex(180,-58){2}
    \Vertex(240,-58){2}
    \Text(55,-53)[lb]{$2a$}
    \Text(115,-53)[lb]{$2b$}
    \Text(163,-61)[lb]{$=$}
    \Text(180,-53)[lb]{$2(a+b)-d$}
    \Text(250,-63)[lb]{$\times U(a,b,d-a-b)$}
  \end{picture}

\end{center}
where the middle point on the l.h.s. is an integrated over vertex.
Additionally, we will make use of the following identity for $a_1+a_2+a_3=d$,
also known as the uniqueness relation
\cite{Parisi:1971,Symanzik:1972wj,Vasiliev:1981yc,Vasiliev:1981dg,Vasiliev:1975mq,Kazakov:1983ns,Kazakov:1984bw}
\begin{equation}
\label{uniqueness}
\int d^dx\,\frac{1}{|x_1-x|^{2a_1}|x_2-x|^{2a_2}|x_3-x|^{2a_3}}
=\frac{U(a_1,a_2,a_3)}{|x_{12}|^{d-2a_3}|x_{13}|^{d-2a_2}|x_{23}|^{d-2a_1}}\,,
\end{equation}
where the function $U$ is defined in (\ref{U definition}).
The uniqueness relation can be represented diagrammatically as \cite{Vasiliev:1981yc,Vasiliev:1981dg,Vasiliev:1975mq}
\begin{center}

  \begin{picture}(210,66) (70,-31)
    \SetWidth{1.0}
    \SetColor{Black}
    \Line[](32,34)(80,2)
    \Line[](80,2)(32,-30)
    \Line[](80,2)(128,2)
    \Line[](192,34)(192,-30)
    \Line[](192,-30)(240,2)
    \Line[](240,2)(192,34)
    \Vertex(32,34){2}
    \Vertex(32,-30){2}
    \Vertex(80,2){4}
    \Vertex(128,2){2}
    \Vertex(192,34){2}
    \Vertex(192,-30){2}
    \Vertex(240,2){2}
    \Text(55,-25)[lb]{$2a_1$}
    \Text(55,22)[lb]{$2a_2$}
    \Text(100,5)[lb]{$2a_3$}
    \Text(160,-1)[lb]{$=$}
    \Text(180,-1)[lb]{$\alpha$}
    \Text(215,-28)[lb]{$\beta$}
    \Text(215,24)[lb]{$\gamma$}
    \Text(250,-5)[lb]{$\times \left(-\frac{2}{\sqrt{N}}\right) U\left(a_1,a_2,a_3\right)$}
  \end{picture}

\end{center}
where the middle vertex on the l.h.s. is assumed to be integrated over, and we introduced
$\alpha = d-2a_3$, $\beta = d-2a_2$, $\gamma = d-2a_1$. Here we also accounted for the Feynman rule associated with the cubic vertex.
\begin{center}

  \begin{picture}(210,66) (70,-31)
    \SetWidth{1.0}
    \SetColor{Black}
    \Line[](32,34)(80,2)
    \Line[](80,2)(32,-30)
    \Line[](80,2)(128,2)
    \Vertex(80,2){4}
    \Text(150,-7)[lb]{$=-\frac{2}{\sqrt{N}}$}
  \end{picture}

\end{center}

\subsection{$\phi$ propagator}
Let us consider the $1/N$ correction to the $\phi$ propagator. It is given by the following 1-loop diagram
\begin{center}

  \begin{picture}(670,60) (125,-40)
    \Text(140,-7)[lb]{$P\;\;=$}
    \Vertex(180,-4){2}
    \Line[](180,-4)(230,-4)
    \Text(175,0)[lb]{$x_1$}
    \Text(215,0)[lb]{$x_3$}
    \Text(285,0)[lb]{$x_4$}
    \Vertex(230,-4){4}
    \Arc[](255,-4)(25,-180,-360)
    \Arc[](255,-4)(25,0,180)
    \Vertex(280,-4){4}
    \Line[](280,-4)(330,-4)
    \Text(325,0)[lb]{$x_2$}
    \Vertex(330,-4){2}
    \Text(195,0)[lb]{\scalebox{0.6}{$2\Delta_\phi$}}
    \Text(253,-42)[lb]{\scalebox{0.6}{$2\Delta_\phi$}}
    \Text(253,25)[lb]{\scalebox{0.6}{$2\Delta_s$}}
    \Text(305,0)[lb]{\scalebox{0.6}{$2\Delta_\phi$}}
  \end{picture}

\end{center}
This diagram reveals a simplest application of the Feynman rules formulated in the previous subsection. It diverges in any $d$, and therefore we regularize it by adding $\delta/2\ll 1$ to the scaling dimension of $s$ \cite{Vasiliev:1981yc,Vasiliev:1981dg,Vasiliev:1975mq}. In other words, we analytically continue the diagram in $\Delta_s$ rather than in $d$. Applying the merging relation (\ref{propagator splitting}) twice, yields
\begin{align}
P(x_1,x_2)&=\left(-\frac{2}{\sqrt{N}}\right)^2 C_\phi^3 C_s\,\mu^{-\delta}\,
\int d^dx_{3,4}\frac{1}{|x_{13}|^{2\Delta_\phi}
|x_{24}|^{2\Delta_\phi}|x_{34}|^{2\Delta_\phi+2\Delta_s+\delta}}\notag\\
&=\frac{4}{N}C_\phi^3 C_s\,\mu^{-\delta}\,
U\left(\Delta_\phi,\Delta_\phi +\Delta_s+\frac{\delta}{2},-\frac{\delta}{2}\right)
\int d^dx_3 \frac{1}{|x_{13}|^{2\Delta_\phi}|x_{23}|^{d+\delta}}\\
&=\frac{4}{N}C_\phi^3 C_s\,\mu^{-\delta}\,
U\left(\Delta_\phi,\Delta_\phi +\Delta_s+\frac{\delta}{2},-\frac{\delta}{2}\right)
U\left(\Delta_\phi,\frac{d+\delta}{2},\frac{d-\delta}{2}-\Delta_\phi\right)\,
\frac{1}{|x_{12}|^{2\Delta_\phi +\delta}}\,.\notag
\end{align}
Combining with the free propagator (\ref{free phi propagator in position space})
and expanding around $\delta=0$, we obtain
\begin{equation}
\label{phi propagator result position space 1}
\langle \phi(x_1)\phi(x_2)\rangle
=\frac{C_\phi}{|x_{12}|^{2\Delta_\phi}}\,\left(1+\left(\frac{2\gamma_\phi}{\delta}
+A_\phi+{\cal O}(\delta)\right)\,\frac{1}{(|x_{12}|\mu)^\delta}\right)\,.
\end{equation}
Here 
\begin{align}
\label{gamma phi 1over N general}
\gamma_\phi &= \frac{1}{N}\,
\frac{2^d \sin \left(\frac{\pi  d}{2}\right) \Gamma \left(\frac{d-1}{2}\right)}
{\pi ^{3/2} (d-2) d  \Gamma \left(\frac{d}{2}-2\right)}\,,\\
\label{A phi 1over N general}
A_\phi &= \gamma_\phi\,\left(\frac{d}{2-d}-\frac{2}{d}\right)\,.
\end{align}
Divergence in the correlation function can be readily removed by the wave function renormalization,
\begin{equation}
\label{general d Z phi}
 \phi = \sqrt{Z_\phi}\,  \tilde \phi\, , \qquad Z_\phi = 1 + \frac{2\gamma_\phi}{\delta}\,.
\end{equation}
As a result, the correlation function for the physical field $\tilde \phi$ to the next-to-leading order in the $1/N$ expansion takes the form (\ref{corrected phi propagator in position space}) with $\gamma_\phi$, $A_\phi$ given by (\ref{gamma phi 1over N general}), (\ref{A phi 1over N general}). 

Note that $\gamma_\phi$ in $d=5$ agrees with (\ref{gamma_phi d=5 result}), whereas in general $d$ our results match \cite{Vasiliev:1981yc,Vasiliev:1981dg,Petkou:1995vu,Petkou:1994ad,Derkachov:1997ch}. In particular, the anomalous dimension (\ref{gamma phi 1over N general}) and the amplitude shift (\ref{A phi 1over N general})
in $d=6-\epsilon$ dimensions are given by
\begin{align}
\gamma_\phi &= \frac{\epsilon}{N}+{\cal O}(\epsilon^2)\,,\\
\label{Aphi 6 - epsilon}
A_\phi &= -\frac{11\epsilon}{6N}+{\cal O}(\epsilon^2)\,.
\end{align}

\subsection{Hubbard-Stratonovich propagator}

Next we derive the $1/N$ correction to the propagator of the auxiliary field $s$. As in section \ref{sec:d5fixed point}, we have three diagrams $C_{1,2,3}$. Furthermore, since the model is regulated by analytic continuation in the scaling dimension of the fields, there is no need to introduce explicitly the counterterm diagrams $C_{4,5}$ of section \ref{sec:d5fixed point}. They are associated with the wave function renormalization and we merely implement it at the end of calculation.

For each diagram we begin by counting and writing down the prefactor associated with the amplitudes $C_{\phi,s}$.
To this end, denote
\begin{equation}
C_1(x_1,x_2) = C_s^3 C_\phi^4\,\tilde C_1(x_1,x_2)\,\mu^{-\delta}\,,
\end{equation}
where $\tilde C_1(x_1,x_2)$ is given by
\begin{center}

  \begin{picture}(194,90) (31,-37)
    \SetWidth{1.0}
    \SetColor{Black}
    \Text(-15,5)[lb]{$\tilde C_1=$}
    \Line[](32,10)(94,10)
    \Text(55,15)[lb]{\scalebox{0.6}{$2\Delta_s$}}
    \Arc[](128,10)(35,153,513)
    \Line[](162,10)(224,10)
    \Text(190,15)[lb]{\scalebox{0.6}{$2\Delta_s$}}
    \Line(128,44)(128,-24)
    \Text(133,15)[lb]{\scalebox{0.6}{$2\Delta_s+\delta$}}
    \Text(73,35)[lb]{\scalebox{0.6}{$2\Delta_\phi-\eta$}}
    \Text(73,-20)[lb]{\scalebox{0.6}{$2\Delta_\phi+\eta$}}
    \Text(157,-20)[lb]{\scalebox{0.6}{$2\Delta_\phi+\eta$}}
    \Text(157,35)[lb]{\scalebox{0.6}{$2\Delta_\phi-\eta$}}
    \Vertex(32,10){2}
    \Vertex(94,10){4}
    \Vertex(162,10){4}
    \Vertex(224,10){2}
    \Vertex(128,-24){4}
    \Vertex(128,44){4}
    \Text(17,5)[lb]{\scalebox{1}{$x_1$}}
    \Text(230,5)[lb]{\scalebox{1}{$x_2$}}
    \Text(80,-5)[lb]{\scalebox{1}{$x_3$}}
    \Text(165,-5)[lb]{\scalebox{1}{$x_6$}}
    \Text(125,50)[lb]{\scalebox{1}{$x_4$}}
    \Text(125,-40)[lb]{\scalebox{1}{$x_5$}}
  \end{picture}

\end{center}
We introduced an extra power $\eta$ to the internal lines representing $\phi$ propagator, see {\it e.g.,}
\cite{Vasiliev:1981yc,Vasiliev:1981dg,Ciuchini:1999wy,Belokurov:1984da}.
Ultimately we are interested in the limit $\eta, \delta\rightarrow 0$. However, this limit
can be reliably taken as long as $\eta={\cal O}(\delta)$. Indeed, the resulting diagram
is both symmetric w.r.t. $\eta\rightarrow-\eta$ and finite as $\eta\to 0$, therefore its series expansion around $\eta=0$ takes the form \cite{Gubser:2017vgc}
\begin{equation}
\tilde C_1= f_0+f_2\,\eta^2+f_4\eta^4+\dots
\end{equation}
Further expanding it around $\delta=0$ and keeping in mind that the coefficients $f_a$
have at most simple poles at $\delta=0$, we conclude that $\tilde C_1=f_0$
in the limit $\delta\rightarrow 0$, as long as $\eta={\cal O}(\delta)$.
It is convenient to choose $\eta=\delta/2$. As a result, we obtain
\begin{align}
\tilde C_1&=\frac{1}{2}\left(-\frac{2}{\sqrt{N}}\right)^4N
\int d^dx_{3,6}\frac{1}{(|x_{13}||x_{26}|)^{2\Delta_s}}
\int d^dx_5\frac{1}{|x_{35}|^{2\Delta_\phi+\frac{\delta}{2}}
|x_{56}|^{2\Delta_\phi+\frac{\delta}{2}}}\notag\\
&\times\int d^dx_4\frac{1}{|x_{34}|^{2\Delta_\phi-\frac{\delta}{2}}
|x_{45}|^{2\Delta_s+\delta}
|x_{46}|^{2\Delta_\phi-\frac{\delta}{2}}}\,.
\end{align}
Presence of $\eta=\delta/2$ makes it possible to use the uniqueness equation (\ref{uniqueness})
to integrate over $x_4$. Applying subsequently the propagator merging relation
(\ref{propagator splitting}) to the integral over $x_5$, yields
\begin{align}
\label{tilde C1 result}
\tilde C_1&=\frac{8}{N}\,
U\left(\Delta_\phi-\frac{\delta}{4},\Delta_\phi-\frac{\delta}{4},
\Delta_s+\frac{\delta}{2}\right)
U\left(\frac{d+\delta}{2},\frac{d+\delta}{2},-\delta \right)\notag\\
&\times\int d^dx_{3,6}\frac{1}{(|x_{13}|
|x_{26}|)^{2\Delta_s}}
\frac{1}{|x_{36}|^{2d-2\Delta_s+\delta}}\,.
\end{align}
We postpone carrying out the remaining integrals over the two edge points $x_{3,6}$. Similar integrals appear in the calculation of $C_{2,3}$, and therefore it makes sense to evaluate them after we have assembled all the terms
$C_{1,2,3}$ together. 

The next diagram contributing to the $s$ propagator at the ${\cal O}(1/N)$ order is denoted by
\begin{equation}
\label{C2 in terms of C tilde 2}
C_2(x_1,x_2) = C_s^4 C_\phi^6 \tilde C_2(x_1,x_2)\,\mu^{-\delta}\,,
\end{equation}
where
\begin{center}

  \begin{picture}(322,100) (31,-37)
    \SetWidth{1.0}
    \SetColor{Black}
    \Text(-15,11)[lb]{$\tilde C_2=$}
    \Text(55,22)[lb]{\scalebox{0.6}{$2\Delta_s$}}
    \Text(320,22)[lb]{\scalebox{0.6}{$2\Delta_s$}}
    \Text(80,-22)[lb]{\scalebox{0.6}{$2\Delta_\phi+\eta$}}
    \Text(275,-22)[lb]{\scalebox{0.6}{$2\Delta_\phi-\eta$}}
    \Text(280,45)[lb]{\scalebox{0.6}{$2\Delta_\phi+\eta$}}
    \Text(80,45)[lb]{\scalebox{0.6}{$2\Delta_\phi-\eta$}}
    \Text(180,55)[lb]{\scalebox{0.6}{$2\Delta_s+\delta/2$}}
    \Text(180,-30)[lb]{\scalebox{0.6}{$2\Delta_s+\delta/2$}}
    \Text(145,11)[lb]{\scalebox{0.6}{$2\Delta_\phi$}}
    \Text(225,11)[lb]{\scalebox{0.6}{$2\Delta_\phi$}}
    \Text(15,11)[lb]{$x_1$}
    \Text(360,11)[lb]{$x_2$}
    \Text(80,5)[lb]{$x_7$}
    \Text(295,5)[lb]{$x_8$}
    \Text(140,55)[lb]{$x_3$}
    \Text(230,55)[lb]{$x_4$}
    \Text(140,-30)[lb]{$x_5$}
    \Text(230,-30)[lb]{$x_6$}
    \Line[](32,16)(94,16)
    \Arc[](128,16)(35,153,513)
    \Line[](144,48)(240,48)
    \Line[](144,-16)(240,-16)
    \Arc[](256,16)(35,153,513)
    \Line[](292,16)(354,16)
    \Vertex(32,16){2}
    \Vertex(94,16){4}
    \Vertex(292,16){4}
    \Vertex(354,16){2}
    \Vertex(144,48){4}
    \Vertex(240,48){4}
    \Vertex(144,-16){4}
    \Vertex(240,-16){4}
  \end{picture}

\end{center}
Notice that we regularized this diagram by supplementing the internal $s$ propagators with an additional power of $\delta/2$ rather than $\delta$. This is done in order to ensure the same total compensating power of $\mu$ on the r.h.s.
of (\ref{C2 in terms of C tilde 2}) as in $C_1$, namely $\mu^{-\delta}$, and therefore such prescription guarantees consistency of regularization of the diagrams \cite{Gubser:2017vgc}. We have also modified some of the $\phi$ lines, relying on the  technique we used in evaluating the diagram $\tilde C_1$. Setting $\eta=\frac{\delta}{2}$  \cite{Ciuchini:1999wy,Belokurov:1984da}, see also \cite{Gubser:2017vgc},
allows us to carry out the integrals over $x_{3,6}$ using the uniqueness relation
(\ref{uniqueness}),
\begin{align}
\tilde C_2&=\frac{1}{2}\,\left(-\frac{2}{\sqrt{N}}\right)^6\,N^2\,
\int d^dx_{7,8}\frac{1}{(|x_{17}||x_{28}|)^{2\Delta_s}}
\int d^dx_{4,5}\frac{1}{|x_{48}|^{2\Delta_\phi+\frac{\delta}{2}}
|x_{57}|^{2\Delta_\phi+\frac{\delta}{2}}}\notag\\
&\times
\int d^dx_3\frac{1}{|x_{37}|^{2\Delta_\phi-\frac{\delta}{2}}
|x_{34}|^{2\Delta_s+\frac{\delta}{2}}
|x_{35}|^{2\Delta_\phi}}
\int d^dx_6\frac{1}{|x_{68}|^{2\Delta_\phi-\frac{\delta}{2}}
|x_{56}|^{2\Delta_s+\frac{\delta}{2}}
|x_{46}|^{2\Delta_\phi}}\,.
\end{align}
Integrating over $x_{3,6}$, gives
\begin{align}
\tilde C_2&=\frac{32}{N}\, U\left(\Delta_\phi-\frac{\delta}{4}, \Delta_s+\frac{\delta}{4}, \Delta_\phi\right)^2 \,
\int d^dx_{7,8}\frac{1}{(|x_{17}||x_{28}|)^{2\Delta_s}}
\tilde c_2(x_7,x_8;0)\,,
\end{align}
where
\begin{align}
\tilde c_2(x_7,x_8;\eta')
=\int d^dx_{4,5}\frac{1}{|x_{48}|^{2d-3\Delta_s-\eta'}
|x_{57}|^{2d-3\Delta_s+\eta'}
|x_{45}|^{2\Delta_s+\delta}
|x_{47}|^{\Delta_s-\eta'}
|x_{58}|^{\Delta_s+\eta'}}\,.
\end{align}
Here we have introduced an additional exponent $\eta'$
\cite{Vasiliev:1981yc,Vasiliev:1981dg,Gubser:2017vgc} which has little impact on the final value of the diagram.
Indeed, by performing the following change of integration variables
\begin{equation}
x_4\rightarrow x_7+x_8-x_5\,,\qquad
x_5\rightarrow x_7+x_8-x_4\,,
\end{equation}
one can see that $\tilde c_2(x_7,x_8;\eta')=\tilde c_2(x_7,x_8;-\eta')$.
Therefore, similarly to the calculation of $\tilde C_1$, any choice of $\eta'={\cal O}(\delta)$ does not change the  value of the diagram in the limit $\delta\rightarrow 0$.
Specifically choosing $\eta'=\frac{\delta}{2}$ makes it possible to apply the uniqueness
relation (\ref{uniqueness}) to the integral over $x_4$ \cite{Gubser:2017vgc} followed by
the propagator merging relation (\ref{propagator splitting})
to the integral over $x_5$
\begin{align}
\label{tilde C2 result}
\tilde C_2&=\frac{32}{N}\,
U\left(\Delta_\phi-\frac{\delta}{4},\Delta_s+\frac{\delta}{4},
\Delta_\phi\right)^2
U\left(d-\frac{3\Delta_s}{2}-\frac{\delta}{4},\Delta_s+\frac{\delta}{2},
\frac{\Delta_s}{2}-\frac{\delta}{4}\right)\notag\\
&\times U\left(\frac{d+\delta}{2},\frac{d+\delta}{2},-\delta \right)\int d^dx_{7,8}\frac{1}{(|x_{17}|
|x_{28}|)^{2\Delta_s}}
\frac{1}{|x_{78}|^{2d-2\Delta_s+\delta}}\,.
\end{align}

Finally, the last diagram is given by
\begin{equation}
C_3(x_1,x_2) = C_s^3 C_\phi^4 \tilde C_3(x_1,x_2)\,\mu^{-\delta}\,,
\end{equation}
where
\begin{center}

  \begin{picture}(194,90) (31,-27)
    \SetWidth{1.0}
    \SetColor{Black}
    \Text(-15,5)[lb]{$\tilde C_3=$}
    \Text(16,5)[lb]{$x_1$}
    \Text(230,5)[lb]{$x_2$}
    \Line[](32,10)(94,10)
    \Text(55,15)[lb]{\scalebox{0.6}{$2\Delta_s$}}
    \Text(78,15)[lb]{\scalebox{0.6}{$2\Delta_\phi$}}
    \Text(168,15)[lb]{\scalebox{0.6}{$2\Delta_\phi$}}
    \Text(80,0)[lb]{$x_3$}
    \Text(168,0)[lb]{$x_4$}
    \Text(85,30)[lb]{$x_5$}
    \Text(163,30)[lb]{$x_6$}
    \Arc[](128,10)(35,153,513)
    \Line[](162,10)(224,10)
    \Text(195,15)[lb]{\scalebox{0.6}{$2\Delta_s$}}
    \Line[](97,25)(159,25)
    \Text(115,15)[lb]{\scalebox{0.6}{$2\Delta_s+\delta$}}
    \Text(125,48)[lb]{\scalebox{0.6}{$2\Delta_\phi$}}
    \Text(125,-36)[lb]{\scalebox{0.6}{$2\Delta_\phi$}}
    \Vertex(32,10){2}
    \Vertex(94,10){4}
    \Vertex(162,10){4}
    \Vertex(224,10){2}
    \Vertex(97,25){4}
    \Vertex(159,25){4}
  \end{picture}

\end{center}
This diagram contains a sub-diagram which was evaluated already, namely,
the one-loop correction to the $\phi$ propagator. Hence, it can be readily simplified
\begin{align}
\label{tilde C3 result}
\tilde C_3&=\frac{16}{N}\,
U\left(\Delta_\phi,\Delta_\phi+\Delta_s+\frac{\delta}{2},
-\frac{\delta}{2}\right)
U\left(\Delta_\phi,\frac{d+\delta}{2},\frac{d-\delta}{2}-\Delta_\phi \right)\notag\\
&\times\int d^dx_{3,4}\frac{1}{(|x_{13}|
|x_{24}|)^{2\Delta_s}}
\frac{1}{|x_{34}|^{2d-2\Delta_s+\delta}}\,.
\end{align}

Next we combine $C_{1,2,3}$.
To begin with, we note that (\ref{tilde C1 result}),
(\ref{tilde C2 result}), (\ref{tilde C3 result}) exhibit a similar structure,
\begin{equation}
\tilde C_i=\hat C_i \int d^dx_{3,4}\frac{1}{(|x_{13}|
|x_{24}|)^{2\Delta_s}}
\frac{1}{|x_{34}|^{2d-2\Delta_s+\delta}}\,.
\end{equation}
Using (\ref{propagator splitting}) twice to perform the remaining integrals over $x_{3,4}$, yields
\begin{equation}
\tilde C_i=U\left(\Delta_s,
d-\Delta_s+\frac{\delta}{2},-\frac{\delta}{2}\right)
U\left(\Delta_s,\frac{d+\delta}{2},
\frac{d-\delta}{2}-\Delta_s\right)\,\hat C_i\,\frac{1}{|x_{12}|^{2\Delta_s + \delta}}\,.
\end{equation}
Summing up all three diagrams and expanding around $\delta=0$ we obtain
\begin{equation}
\label{s propagator general result}
\langle s(x_1)s(x_2)\rangle
=\frac{C_s}{|x_{12}|^{2\Delta_s}}\,\left(
1+\left(\frac{2\gamma_s}{\delta}+A_s+{\cal O}(\delta)\right)\,\frac{1}{(|x_{12}\mu|)^\delta}\right)\,,
\end{equation}
with
\begin{align}
\label{gamma s result}
\gamma_s &= \frac{1}{N}\,\frac{4 \sin \left(\frac{\pi  d}{2}\right) \Gamma (d)}
{\pi  \Gamma \left(\frac{d}{2}+1\right) \Gamma \left(\frac{d}{2}-1\right)}\,,\\
\label{A s 1over N general}
A_s&=2\gamma_\phi\,\left(\frac{d(d-3)+4}{4-d}
\left( H_{d-3}+\pi  \cot \Big(\frac{\pi  d}{2}\Big)\right)
+\frac{8}{(d-4)^2}+\frac{2}{d-2}+\frac{2}{d}-2 d-1\right)\,,
\end{align}
where $H_{n}$ is the $n$\textit{th} harmonic number.

As before, divergence in the correlation function is removed by the wave function renormalization,
\begin{equation}
\label{general d Z s}
s=\sqrt{Z_s} \tilde s \, , \qquad Z_s = 1 + \frac{2\gamma_s}{\delta}\,.
\end{equation}
In particular, the correlation function for the physical field $\tilde s$ to the next-to-leading order in the $1/N$ expansion takes the form (\ref{corrected s propagator in position space}) with $\gamma_s$, $A_s$ given by  (\ref{gamma s result}), (\ref{A s 1over N general}).

Notice that the wave function renormalization constants (\ref{general d Z phi}), (\ref{general d Z s}) generate
the following renormalization of the bare term in the action (\ref{main action})
\begin{equation}
\frac{1}{\sqrt{N}}\,\phi ^ 2 s = \frac{1}{\sqrt{N}}\,\tilde\phi ^ 2 \tilde s
+ \frac{2\gamma_\phi + \gamma_s}{\delta} \, \frac{1}{\sqrt{N}}\,\tilde\phi ^ 2 \tilde s\,,
\end{equation}
where we keep $\mathcal{O}(1/N)$ corrections only. Thus the counterterm is given by
\begin{equation}
\label{S int ct in general dimension}
S_{\textrm{int}}^{\textrm{c.t.}} = \frac{2\gamma_\phi + \gamma_s}{\delta} \, \frac{1}{\sqrt{N}}\,\tilde\phi ^ 2 \tilde s\,.
\end{equation}
To avoid clutter we suppress tilde above the renormalized fields $\tilde\phi$, $\tilde s$ in what follows. Of course,
 physical correlations are associated with the renormalized fields.

Before closing this section we note that $\gamma_s$ in $d=5$ agrees with (\ref{gamma_s d=5 result}), whereas in general $d$ our results match \cite{Vasiliev:1981yc,Vasiliev:1981dg,Petkou:1995vu,Petkou:1994ad,Derkachov:1997ch}. In particular, the anomalous dimension (\ref{gamma s result}) and the amplitude shift (\ref{A s 1over N general}) in $d=6-\epsilon$ dimensions are given by
\begin{align}
\gamma_s &= \frac{40\epsilon}{N}+{\cal O}(\epsilon^2)\,,\\
\label{As 6 - epsilon}
A_s &= \frac{44}{N}+{\cal O}(\epsilon)\,.
\end{align}

\section{Conformal three-point functions}
\label{sec:three-point}

In this section we calculate the next-to-leading order correction to
the OPE coefficients of the conformal three-point functions 
$\langle \phi \phi s\rangle$ and $\langle s s s \rangle$ associated with the fundamental
scalar field $\phi$ and the auxiliary Hubbard-Stratonovich field $s$. In a general $d$-dimensional
$O(N)$ symmetric CFT,
the results for $\langle \phi \phi s\rangle$ to the next-to-leading order
and leading order $\langle sss\rangle$ correlator
are known based on the conformal
bootstrap approach \cite{Petkou:1995vu,Petkou:1994ad}
(see also \cite{Lang:1993ct,Leonhardt:2003du}
for discussion in the context of $O(N)$ sigma model). In contrast, we recover these results 
in the critical $\phi^4$ model from the new perspective which has not previously been given in the literature, and extend the calculations to derive a new OPE coefficient of the three-point function $\langle sss\rangle$. 

Furthermore, to facilitate and make progress in the calculations we evaluate the 4-loop conformal triangle diagram, and the associated 3-loop trapezoid graph, as well as the 3-loop bellows diagram in general dimension $d$. To the best of our knowledge, these diagrams were not evaluated in the literature before, and the corresponding analytic expressions were not displayed elsewhere. These diagrams naturally appear in numerous conformal calculations, and therefore might be useful in the future.

\subsection{$\langle \phi \phi s\rangle$}
\label{sec:phi phi s}

The leading order result for the $\langle \phi \phi s\rangle$ correlator follows directly from the tree-level diagram,
\begin{equation}
\langle \phi(x_1) \phi(x_2) s(x_3)\rangle \Bigg|_{\textrm{leading}} 
=-\frac{2}{\sqrt{N}}\,C_\phi^2 C_s\,\int d^dx\frac{1}{|x_1-x|^{2\Delta_\phi}
|x_2-x|^{2\Delta_\phi}|x_3-x|^{2\Delta_s}}\,.
\end{equation}
Using the uniqueness relation (\ref{uniqueness}) we obtain
the conformal three-point function
\begin{equation}
\langle \phi(x_1) \phi(x_2) s(x_3)\rangle \Bigg|_{\textrm{leading}} 
=\frac{C_{\phi\phi s}}{|x_{12}|^{2\Delta_\phi-\Delta_s}|x_{13}|^{\Delta_s}
|x_{23}|^{\Delta_s}}\,,
\end{equation}
where the leading order coefficient is given by
\begin{equation}
\label{Original phi phi s}
C_{\phi\phi s}=-\frac{2}{\sqrt{N}}\,
C_\phi^2 C_sU(\Delta_\phi, \Delta_\phi, \Delta_s) \,.
\end{equation}
It is convenient to express the three-point functions in terms of normalized fields
\begin{equation}
\label{field normalizations}
\phi\rightarrow \phi \, \sqrt{C_\phi(1+A_\phi)}\,,\qquad 
s\rightarrow s \, \sqrt{C_s(1+A_s)}\,,
\end{equation}
\textit{i.e.}, fields whose two-point functions are normalized to unity. The OPE coefficient, $\tilde C_{\phi\phi s}$, associated with the renormalized fields is given
by\footnote{We ignore $A_{\phi,s}$ in (\ref{field normalizations}) to leading order in the $1/N$ expansion.}
\begin{equation}
\tilde C_{\phi\phi s} = \frac{C_{\phi\phi s}}{C_\phi C_s^\frac{1}{2}}\,,
\end{equation}
or equivalently  \cite{Petkou:1995vu,Petkou:1994ad},
\begin{equation}
\label{phi phi s leading}
\tilde C_{\phi\phi s} =-\frac{2}{\sqrt{N}}\, C_\phi C_s^\frac{1}{2}U(\Delta_\phi, \Delta_\phi, \Delta_s) \,.
\end{equation}
In particular, in $d=6-\epsilon$ dimensions we obtain
\begin{equation}
\tilde C_{\phi\phi s} = - \sqrt{\frac{6\epsilon}{N}} + {\cal O}(\epsilon)\,.
\end{equation}

Next we evaluate the $1/N$ correction to the three-point coefficient $\tilde C_{\phi\phi s}$ in general dimension,
and demonstrate that it agrees with \cite{Petkou:1995vu,Petkou:1994ad}.\footnote{
We thank Simone Giombi, Igor Klebanov and Gregory Tarnopolsky for attracting our attention to \cite{Petkou:1994ad} and encouraging us to carry out the calculation in general dimension.}
In fact, the only known complete derivation of the $\langle\phi\phi s\rangle$ three-point
function to the next-to-leading order in the large-$N$ expansion has been given
in  \cite{Petkou:1995vu,Petkou:1994ad} by solving the consistency constraints in the  bootstrap approach. 
Therefore a direct diagrammatic technique, which is followed below, provides an independent derivation of the result for the $\langle\phi\phi s\rangle$ three-point function.

We begin by determining the dressed propagators and vertices,
which will also be used later to calculate the next-to-leading order correction to the 
three-point function $\langle sss\rangle$.
The relevant diagrams
are obtained by dressing the propagators and $\phi\phi s$ interaction
vertex of the tree level diagram. Let us denote the dressed propagators
(\ref{corrected phi propagator in position space}) and 
(\ref{corrected s propagator in position space}) by a line and a grey blob,
\begin{center}

  \begin{picture}(162,34) (130,-26)
    \SetWidth{1.0}
    \SetColor{Black}
    \Line[](144,-14)(208,-14)
    \Text(115,5)[lb]{\scalebox{0.6}{$2(\Delta+\gamma)$}}
    \GOval(128,-14)(16,16)(0){0.882}
    \Line[](48,-14)(112,-14)
    \Vertex(48,-14){2}
    \Vertex(208,-14){2}
    \Text(215,-17)[lb]{$x_2$}
    \Text(32,-17)[lb]{$x_1$}
    \Text(240,-23)[lb]{$=C (1+A)\,\frac{\mu^{-2\gamma}}
{|x_{12}|^{2(\Delta + \gamma)}}$}
  \end{picture}

\end{center}
Here the set of parameters $(C,A,\Delta,\gamma)$ stands either for
$(C_\phi,A_\phi,\Delta_\phi,\gamma_\phi)$, when the $\phi$-propagator is considered, or
for $(C_s,A_s,\Delta_s,\gamma_s)$, when the line represents $s$-propagator.

The full three-point function can then be obtained by evaluating the following diagram
\begin{center}

  \begin{picture}(260,182) (31,-47)
    \SetWidth{1.0}
    \SetColor{Black}
    \Line[](96,34)(128,34)
    \GOval(144,34)(16,16)(0){0.882}
    \Line[](180,70)(156,46)
    \Line[](180,-2)(154,22)
    \GOval(192,82)(16,16)(0){0.882}
    \Line[](228,118)(204,94)
    \Vertex(228,118){2}
    \Text(235,118)[lb]{$x_1$}
    \Text(235,-55)[lb]{$x_2$}
    \GOval(192,-14)(16,16)(0){0.882}
    \Line[](228,-50)(204,-26)
    \Vertex(228,-50){2}
    \Text(65,55)[lb]{\scalebox{0.6}{$2(\Delta_s+\gamma_s)$}}
    \Text(30,20)[lb]{$x_3$}
    \GOval(80,34)(16,16)(0){0.882}
    \Line[](32,34)(64,34)
    \Vertex(32,34){2}
    \Text(170,102)[lb]{\scalebox{0.6}{$2(\Delta_\phi+\gamma_\phi)$}}
    \Text(170,-40)[lb]{\scalebox{0.6}{$2(\Delta_\phi+\gamma_\phi)$}}
  \end{picture}
\end{center}
where the grey blob in the center represents a dressed three-point vertex.\footnote{By definition, the dressed vertex has no propagators attached to the external legs (amputated diagram).} By conformal invariance this diagram takes the form
\begin{equation}
\label{s phi phi expected form}
\langle \phi(x_1)\phi(x_2)s(x_3)\rangle =\tilde C_{\phi\phi s}\,\frac{(1+W_{\phi\phi s})\mu^{-2\gamma_\phi-\gamma_s}}
{|x_{12}|^{2(\Delta_\phi+\gamma_\phi)-(\Delta_s+\gamma_s)}
|x_{13}|^{\Delta_s+\gamma_s}|x_{23}|^{\Delta_s+\gamma_s}}\,,
\end{equation}
where the fields $\phi$ and $s$ are normalized such that their two-point function has unit amplitude. Our goal is to calculate $W_{\phi\phi s}$ to order ${\cal O}\left(\frac{1}{N}\right)$. It represents a subleading correction to the OPE coefficient $\tilde C_{\phi\phi s}$. 

The dressed 1PI vertex, $\Gamma(x,y,z)$, in the above diagram was evaluated in \cite{Derkachov:1997ch} to $\mathcal{O}(1/N)$ order, see eq. (4.18) there,
\begin{align}
\label{conformal triangle phi phi s integral}
\int\int\int \Gamma(x_1,x_2,x_3) \phi(x_1)\phi(x_2)s(x_3)= \frac{\hat Z}{\sqrt{N}}\int\int\int
\frac{ \mu^{-2\gamma_\phi-\gamma_s} \phi(x_1)\phi(x_2)s(x_3)}{|x_3-x_1|^{2\alpha}|x_3-x_2|^{2\alpha}|x_1-x_2|^{2\beta}}\,,
\end{align}
where in notations of \cite{Derkachov:1997ch}
\begin{equation}
\alpha = \Delta_\phi-\frac{\gamma_s}{2}\,,\quad
\beta = \Delta _ s - \gamma_\phi +\frac{\gamma_s}{2}\,,
\end{equation}
and $\hat Z$ is a constant given by eq. (A.4) in \cite{Derkachov:1997ch} 
\footnote{Our notations differ from those of \cite{Derkachov:1997ch}.
The precise relation is $\mu\rightarrow\tilde\mu$, $\gamma_\psi\rightarrow\gamma_s$,
$H(\Delta)\rightarrow A(\Delta)$.}
\begin{align}
&\hat Z = -\frac{\chi}{2}\,\frac{A(1)^2A(\tilde \mu-2)\Gamma(\tilde \mu)}{\pi^{2\tilde\mu}}\,
\left(1+\delta V'\right)\,,\quad \chi = - (2\gamma_\phi+\gamma_s)\,,\notag \\
&A(\Delta) = \frac{\Gamma\left(\frac{d}{2}-\Delta\right)}{\Gamma(\Delta)} \,,\quad \tilde\mu = \frac{d}{2}\,,\\
& \delta V' = \frac{\eta_1}{N}\frac{ (\tilde\mu -3) \left(6 \tilde\mu ^2-9 \tilde\mu +2\right)}{(\tilde\mu -2)^2}\,,\quad
\eta_1 \equiv \frac{4 (2-\tilde\mu ) \Gamma (2 \tilde\mu -2)}{\Gamma (\tilde\mu -1)^2 \Gamma (2-\tilde\mu ) 
\Gamma (\tilde\mu +1)}\,. \notag
\end{align}

To get $W_{\phi\phi s}$ one should simply attach the dressed propagators to $\Gamma(x,y,z)$ and apply the uniqueness formula (\ref{uniqueness}) three times, because all 3 vertices $x_{1,2,3}$ happen to be unique.  Normalizing the external legs, yields
\begin{equation}
\label{conformal triangle phi phi s integrated}
\tilde C_{\phi\phi s} \, (1 + W_{\phi\phi s}) =-\frac{2}{\sqrt{N}}\frac{C_\phi^2 (1+A_\phi)^2 C_s(1+A_s)}{\sqrt{C_\phi^2 (1+A_\phi)^2C_s(1+A_s)}}\hat Z \hat U\,,
\end{equation}
where we defined
\begin{align}
\hat U &= U\left(\Delta_\phi-\frac{\gamma_s}{2},
\Delta_\phi-\frac{\gamma_s}{2},
\Delta_s+\gamma_s\right)
U\left(\Delta_\phi-\frac{\gamma_s}{2},
\Delta_\phi+\gamma_\phi,
\Delta_s-\gamma_\phi+\frac{\gamma_s}{2}\right)\notag\\
&\times U\left(\Delta_\phi+\gamma_\phi,
\frac{\Delta_s+\gamma_s}{2},
\frac{d}{2}-\gamma_\phi-\frac{\gamma_s}{2}\right)\,.
\label{hat U definition}
\end{align}
Expansion of $\hat U$ in $1/N$ has the following form
\begin{equation}
\label{hat U expansion}
\hat U = N \, u_1 +  u_2 +{\cal O}\left(\frac{1}{N}\right)\,.
\end{equation}
Notice that it starts with the divergent
in $N\rightarrow \infty$ piece $N\,u_1$, originating from the singular in that limit factor
$U\left(\frac{d}{2}-\gamma_\phi-\frac{\gamma_s}{2},\cdots\right)$ in  
(\ref{hat U definition}).
It is convenient to denote the ratio
of the sub-leading to leading coefficients in (\ref{hat U expansion}) by
\begin{equation}
\hat u = \frac{u_2}{N u_1} = 
\frac{\gamma _s ((26-3 d) d-44)+2 \gamma_ \phi  ((d-6) d+4)}{2 (d-4) (d-2) }\,.
\end{equation}
Here the anomalous dimensions $\gamma_\phi$, $\gamma_s$
are given by (\ref{gamma phi 1over N general}), (\ref{gamma s result}).
Combining $\hat u$ with the other next-to-leading order terms in (\ref{conformal triangle phi phi s integrated}), we obtain
\begin{equation}
\label{general W phi phi s}
W_{\phi\phi s} = A_\phi +\frac{A_s}{2}+\delta V\,,
\end{equation}
where
\begin{equation}
\delta V = \hat u + \delta V' 
= \frac{1}{N} \, \frac{2^{d-3} (d (d (5 d-42)+116)-96) \sin \left(\frac{\pi  d}{2}\right) \Gamma \left(\frac{d-1}{2}\right)}{\pi ^{3/2} (d-4) (d-2) \Gamma \left(\frac{d}{2}+1\right)}\,,
\end{equation}
which we can re-write as
\begin{equation}
\label{delta V in general d}
\delta V =\gamma_\phi\, \left(\frac{8}{(d-4)^2}+\frac{2}{d-2}+\frac{6}{d-4}+5\right)\,,
\end{equation}
where $\gamma_\phi$ is given by (\ref{gamma phi 1over N general}).
Substituting (\ref{A phi 1over N general}), (\ref{A s 1over N general}) and (\ref{delta V in general d}) into 
(\ref{general W phi phi s}) gives $W_{\phi\phi s}$ in general dimension. It matches \cite{Petkou:1995vu,Petkou:1994ad} where the generic $O(N)$ symmetric CFT was studied, {\it e.g.,} see eq. (21) in \cite{Petkou:1995vu}.\footnote{Note that the leading terms
in the large $N$ expansion on the l.h.s. and r.h.s. of (\ref{conformal triangle phi phi s integrated})
agree, as one can explicitly verify by direct calculation,
\begin{equation}
-\frac{2}{\sqrt{N}}\,C_\phi C_s^\frac{1}{2}\,
 \left(-\frac{\chi}{2}\,\frac{A(1)^2A(\tilde \mu-2)\Gamma(\tilde \mu)}{\pi^{2\tilde\mu}}\right)\,N\,u_1=
-\frac{2}{\sqrt{N}}C_\phi C_s^\frac{1}{2}U(\Delta_\phi,\Delta_\phi,\Delta_s)=\tilde C_{\phi\phi s}\,.
\end{equation}
} In particular, the above $W_{\phi\phi s}$ agrees with its counterpart in a CFT emerging at the IR fixed point of the $O(N)$ vector model with cubic interactions in $d=6-\epsilon$ dimensions \cite{Fei:2014yja}. In this case one gets
\begin{equation}
\label{d6 W phi phi s}
\delta V= \frac{1}{N}\,\frac{21\epsilon}{2}+{\cal O}(\epsilon^2)\,, \quad
W_{\phi\phi s} = \frac{22}{N} + {\cal O}(\epsilon)\,.
\end{equation}

While this way of calculating $W_{\phi\phi s}$ is straightforward, it does not separate the effect of the anomalous dimensions intrinsic to the propagators from the contribution of the dressed vertex. Such a separation proves to be useful when we evaluate $\langle sss\rangle$ correlation function. In other words, the impact of anomalous dimensions inherent to the propagators is singled out in our method of calculating the $\langle sss \rangle$ correlation function. Hence, for future purpose we define two additional Feynman rules.

To begin with, we exclude $A_{\phi, s}$ from the residue of the dressed propagators. These constants represent $1/N$ corrections to the amplitudes of the two-point functions and can be accounted at the very end.  The dressed propagators without $A_{\phi, s}$ will be denoted by a solid black blob\footnote{Note that it is crucial to account for $A_{\phi, s}$, see {\it e.g.,} (\ref{general W phi phi s}).}
\begin{center}
  \begin{picture}(147,31) (47,-33)
    \SetWidth{1.0}
    \SetColor{Black}
    \Line[](50,-19)(150,-19)
    \Text(159,-29)[lb]{$=C\,\frac{\mu^{-2\gamma}}{|x|^{2(\Delta+\gamma)}}$}
    \Vertex(103,-19){8}
    \Vertex(50,-19){2}
    \Vertex(150,-19){2}
  \end{picture}
\end{center}
As before $(C,\Delta,\gamma)$ stand for $(C_{\phi, s},\Delta_{\phi, s},\gamma_{\phi, s})$,
depending on the considered propagator.

Next we define a dressed vertex with the bare propagators 
attached to it, {\it i.e.,} propagators without anomalous dimensions and $A_{\phi, s}$. Diagrammatically
it is given by
\begin{center}
\begin{equation}
  \begin{picture}(590,70) (17,-20)
    \SetWidth{1.0}
    \SetColor{Black}
    \scalebox{0.75}{
    \GOval(84,28)(18,18)(0){0.882}
    \Line[](96,40)(126,64)
    \Line[](126,-8)(96,16)
    \Line[](66,28)(24,28)
    \Vertex(24,28){2}
    \Vertex(126,64){2}
    \Vertex(126,-8){2}
    \Vertex(222,28){2}
    \Vertex(324,64){2}
    \Vertex(324,-8){2}
    \Line[](222,28)(324,64)
    \Line[](324,64)(324,-8)
    \Line[](324,-8)(222,28)
    \Text(143,23)[lb]{\scalebox{0.9}{$=C_{\phi\phi s}(1+\delta V)$}}
    \Text(100,55)[lb]{\scalebox{0.6}{$2\Delta_\phi$}}
    \Text(100,-5)[lb]{\scalebox{0.6}{$2\Delta_\phi$}}
    \Text(40,32)[lb]{\scalebox{0.6}{$2\Delta_s$}}
    \Text(253,25)[lb]{\scalebox{0.6}{$2(\Delta_\phi+\gamma_\phi)-(\Delta_s+\gamma_s)$}}
    \Text(252,53)[lb]{\scalebox{0.6}{$\Delta_s+\gamma_s$}}
    \Text(252,1)[lb]{\scalebox{0.6}{$\Delta_s+\gamma_s$}}
    \Vertex(366,28){2}
    \Vertex(462,-8){2}
    \Vertex(462,64){2}
    \Line[](366,28)(420,28)
    \Line[](420,28)(462,64)
    \Line[](420,28)(462,-8)
    \Vertex(420,28){4}
    \Line[](504,28)(558,28)
    \Vertex(504,28){2}
    \Vertex(558,28){4}
    \Line[](558,28)(600,64)
    \Line[](558,28)(600,-8)
    \Vertex(600,-8){2}
    \Vertex(600,64){2}
    \Vertex(390,28){8}
    \Vertex(440,46){8}
    \Vertex(440,10){8}
    \Text(340,26)[lb]{\scalebox{0.6}{$-$}}
    \Text(480,26)[lb]{\scalebox{0.6}{$+$}}
    
    \Text(425,60)[lb]{\scalebox{0.6}{$2(\Delta_\phi+\gamma_\phi)$}}
    \Text(425,-12)[lb]{\scalebox{0.6}{$2(\Delta_\phi+\gamma_\phi)$}}
    \Text(375,40)[lb]{\scalebox{0.6}{$2(\Delta_s+\gamma_s)$}}
    
    \Text(575,55)[lb]{\scalebox{0.6}{$2\Delta_\phi$}}
    \Text(575,-5)[lb]{\scalebox{0.6}{$2\Delta_\phi$}}
    \Text(525,32)[lb]{\scalebox{0.6}{$2\Delta_s$}}
    }
  \end{picture}
    \label{Dressed vertex diagram}
\end{equation}
\end{center}
where the external legs are not normalized, and therefore $C_{\phi\phi s}$ rather than $\tilde C_{\phi\phi s}$ appears on the r.h.s. The term $\delta V \sim \mathcal{O}(1/N)$ equals $W_{\phi\phi s}$ up to corrections $A_{\phi, s}$ that we stripped off. In fact, $\delta V$ is closely related to the sum of two loop diagrams, which have been studied
in momentum space in section \ref{sec:interaction vertex in d=5}.
Notice that $\delta V$ is rendered finite, because besides those two loop diagrams there is a contribution from the  vertex counterterm (\ref{S int ct in general dimension}). The last diagram on the r.h.s. simply subtracts the leading $1/N$ contribution from the second diagram
\begin{center}
  \begin{picture}(590,70) (17,-20)
    \SetWidth{1.0}
    \SetColor{Black}
    \scalebox{0.75}{
    \Vertex(16,28){2}
    \Vertex(112,-8){2}
    \Vertex(112,64){2}
    \Line[](16,28)(70,28)
    \Line[](70,28)(112,64)
    \Line[](70,28)(112,-8)
    \Vertex(70,28){4}
    \Line[](154,28)(208,28)
    \Vertex(154,28){2}
    \Vertex(208,28){4}
    \Line[](208,28)(250,64)
    \Line[](208,28)(250,-8)
    \Vertex(250,-8){2}
    \Vertex(250,64){2}
    \Vertex(40,28){8}
    \Vertex(90,46){8}
    \Vertex(90,10){8}
    \Text(130,26)[lb]{\scalebox{1}{$-$}}
    \Text(250,17)[lb]{\scalebox{1}{$= -\frac{2}{\sqrt{N}}\,C_\phi^2 C_s
    \int d^dx_4\,\frac{1}{(|x_{14}||x_{24}|)^{2\Delta_\phi}|x_{34}|^{2\Delta_s}}\,
    \left( \frac{2}{\delta\mu^\delta}\,
    \left(\frac{\gamma_\phi}{|x_{14}|^\delta}+\frac{\gamma_\phi}{|x_{24}|^\delta}
    +\frac{\gamma_s}{|x_{34}|^\delta}\right)-2\frac{2\gamma_\phi+\gamma_s}{\delta}\right)$}}   
    
    \Text(75,60)[lb]{\scalebox{0.6}{$2(\Delta_\phi+\gamma_\phi)$}}
    \Text(75,-12)[lb]{\scalebox{0.6}{$2(\Delta_\phi+\gamma_\phi)$}}
    \Text(25,40)[lb]{\scalebox{0.6}{$2(\Delta_s+\gamma_s)$}}
    
    \Text(225,55)[lb]{\scalebox{0.6}{$2\Delta_\phi$}}
    \Text(225,-5)[lb]{\scalebox{0.6}{$2\Delta_\phi$}}
    \Text(175,32)[lb]{\scalebox{0.6}{$2\Delta_s$}}
    }
  \end{picture}
\end{center}
where the last term within the parenthesis stands for the counterterm associated with the wave function renormalization (\ref{general d Z phi}) and (\ref{general d Z s}), and we used regularized expressions for the dressed propagators obtained previously by the direct calculation of the Feynman graphs. Note that the newly defined vertex with the bare propagators attached to it has logarithmic terms proportional to the anomalous dimensions. Such terms emerge because of renormalization of the vertex, and are intrinsic to the vertex itself rather than to the propagators. 

We choose to keep track separately of the contributions associated with the dressed vertex correction $\delta V$ and anomalous dimensions $\gamma_{\phi,s}$ of the propagators. Hence, the above diagrams will be used extensively in what follows. This approach turns out to be useful when we evaluate the $\langle sss \rangle$ correlation function.

\subsection{$\langle sss\rangle$}

In this subsection we calculate the three-point function $\langle sss \rangle$ to
 ${\cal O}\left(\frac{1}{N^{3/2}}\right)$ order. For the normalized field
(\ref{field normalizations}) it has the form
\begin{align}
\label{sss general}
\langle s(x_1)s(x_2)s(x_3)\rangle =
\tilde C_{s^3}\,\frac{(1+W_{s^3})\mu^{-3\gamma_s}}
{|x_{12}|^{\Delta_s+\gamma_s}
|x_{13}|^{\Delta_s+\gamma_s}|x_{23}|^{\Delta_s+\gamma_s}}\,.
\end{align}
We are going to calculate the leading coefficient $\tilde C_{s^3}$ and the $1/N$ correction
$W_{s^3}$.

The leading behaviour is completely determined by the one-loop triangle diagram
\begin{center}

  \begin{picture}(162,162) (15,-31)
    \SetWidth{1.0}
    \SetColor{Black}
    \Line[](96,130)(96,66)
    \Line[](96,66)(64,18)
    \Line[](96,66)(128,18)
    \Line[](64,18)(128,18)
    \Line[](128,18)(182,-20)
    \Line[](64,18)(9,-20)
    \Vertex(182,-20){2}
    \Vertex(9,-20){2}
    \Vertex(128,18){4}
    \Vertex(64,18){4}
    \Vertex(96,130){2}
    \Vertex(96,66){4}
    \Text(92,21)[lb]{\scalebox{0.6}{$2\Delta_\phi$}}
    \Text(66,43)[lb]{\scalebox{0.6}{$2\Delta_\phi$}}
    \Text(116,43)[lb]{\scalebox{0.6}{$2\Delta_\phi$}}
    \Text(100,95)[lb]{\scalebox{0.6}{$2\Delta_s$}}
    \Text(20,0)[lb]{\scalebox{0.6}{$2\Delta_s$}}
    \Text(158,0)[lb]{\scalebox{0.6}{$2\Delta_s$}}
    \Text(103,125)[lb]{$x_1$}
    \Text(103,65)[lb]{$x_6$}
    \Text(47,17)[lb]{$x_4$}
    \Text(135,17)[lb]{$x_5$}
    \Text(-7,-22)[lb]{$x_2$}
    \Text(188,-22)[lb]{$x_3$}
  \end{picture}

\end{center}
Normalizing the external legs yields
\begin{align}
\langle s(x_1)s(x_2)s(x_3)\rangle\Bigg|_{\textrm{leading}} &=
N\left(-\frac{2}{\sqrt{N}}\right)^3\frac{C_\phi^3 C_s^3}{\sqrt{C_s^3}}
\int d^dx_{4,5}\frac{1}{|x_{35}|^{2\Delta_s}
|x_{45}|^{2\Delta_\phi}|x_{24}|^{2\Delta_s}}\notag\\
&\times \int d^dx_6\frac{1}{|x_{16}|^{2\Delta_s}|x_{46}|^{2\Delta_\phi}
|x_{56}|^{2\Delta_\phi}}\,.
\end{align}
Applying uniqueness formula (\ref{uniqueness}) to the integrals over $x_{4,5,6}$,
we arrive at the leading $\langle sss\rangle$ triangle expression
\begin{equation}
\langle s(x_1)s(x_2)s(x_3)\rangle\Bigg|_{\textrm{leading}}=
\frac{\tilde C_{s^3}}{|x_{12}|^{\Delta_s}|x_{23}|^{\Delta_s}|x_{13}|^{\Delta_s}}\,,
\end{equation}
where
\begin{equation}
\label{Cs3 answer}
\tilde C_{s^3}=-\frac{8}{\sqrt{N}}\,C_\phi^3 C_s^\frac{3}{2}\,
U(\Delta_\phi, \Delta_\phi, \Delta_s)^2 \, U\left(\frac{\Delta_s}{2},
\Delta_s,d-\frac{3\Delta_s}{2}\right)\,.
\end{equation}
One can readily show that for general $d$
\begin{equation}
\label{relation 1}
C_\phi^2 C_s\,
U(\Delta_\phi, \Delta_\phi, \Delta_s) U\left(\frac{\Delta_s}{2},
\Delta_s,d-\frac{3\Delta_s}{2}\right)=\frac{d-3}{2}\,,
\end{equation}
and therefore using (\ref{phi phi s leading}) we obtain
\begin{equation}
\label{Cs3 and Cphiphi s relation}
\tilde C_{s^3} = 2(d-3)\,\tilde C_{\phi\phi s}\,,
\end{equation}
in agreement with \cite{Petkou:1995vu,Petkou:1994ad}.

Next we evaluate the sub-leading term $W_{s^3}$
in (\ref{sss general}). It is determined by the ${\cal O}(1/N^{3/2})$ diagrams with three external legs of type $s$. A large portion of them is obtained by dressing the constituents of the leading order triangle diagram with $1/N$ corrections. Before studying this class of diagrams, recall  that the dressed $\phi$- and $s$-propagators have a non-trivial $1/N$ correction due to the amplitudes $A_\phi$, $A_s$, as well as due to the anomalous dimensions $\gamma_\phi$, $\gamma_s$. We will account for the contribution related to $A_{\phi, s}$ later in this subsection, while for now we focus on studying the effect associated with the dressed vertex (\ref{Dressed vertex diagram}) and the anomalous dimensions $\gamma_{\phi, s}$ inherent to the dressed propagators.

Quite surprisingly, it turns out that various diagrams obtained by dressing the leading $\langle sss\rangle$ triangle can be grouped in such a way that the integrals over their internal vertices can be carried out using the uniqueness relation. We illustrate how it works now.

Dressing each of the three $\phi \phi s$ vertices of the leading $\langle sss\rangle$ triangle diagram gives
\begin{center}
  \begin{picture}(400,162) (15,-31)
    \SetWidth{1.0}
    \SetColor{Black}
    \Line[](96,130)(96,66)
    \Line[](96,66)(64,18)
    \Line[](96,66)(128,18)
    \Line[](64,18)(128,18)
    \Line[](128,18)(182,-20)
    \Line[](64,18)(9,-20)
    \Vertex(182,-20){2}
    \Vertex(9,-20){2}
    \Vertex(128,18){4}
    \Vertex(64,18){4}
    \Vertex(96,130){2}
    \GOval(96,66)(16,16)(0){0.882}
    \Text(92,21)[lb]{\scalebox{0.6}{$2\Delta_\phi$}}
    \Text(66,43)[lb]{\scalebox{0.6}{$2\Delta_\phi$}}
    \Text(116,43)[lb]{\scalebox{0.6}{$2\Delta_\phi$}}
    \Text(100,95)[lb]{\scalebox{0.6}{$2\Delta_s$}}
    \Text(20,0)[lb]{\scalebox{0.6}{$2\Delta_s$}}
    \Text(158,0)[lb]{\scalebox{0.6}{$2\Delta_s$}}
    \Text(103,125)[lb]{$x_1$}
    \Text(47,17)[lb]{$x_4$}
    \Text(135,17)[lb]{$x_5$}
    \Text(-7,-22)[lb]{$x_2$}
    \Text(188,-22)[lb]{$x_3$}
    \Text(208,43)[lb]{\scalebox{1}{$+\quad$ two cyclic permutations $x_1\rightarrow x_2\rightarrow x_3\rightarrow x_1$.}} 
  \end{picture}
\end{center}
Here the vertex blob is determined by (\ref{Dressed vertex diagram}).
In addition, each of the three internal $\phi$-propagators and each of the three
external $s$-propagators in the leading triangle diagram need to be endowed with the anomalous dimensions. The corresponding diagram is given by
\begin{center}
  \begin{picture}(162,162) (15,-31)
    \SetWidth{1.0}
    \SetColor{Black}
    \Line[](96,130)(96,66)
    \Vertex(96,98){8}
    \Line[](96,66)(64,18)
    \Vertex(80,42){8}
    \Line[](96,66)(128,18)
    \Vertex(112,42){8}
    \Line[](64,18)(128,18)
    \Vertex(96,18){8}
    \Line[](128,18)(182,-20)
    \Vertex(155,-1){8}
    \Line[](64,18)(9,-20)
    \Vertex(36,-1){8}
    \Vertex(182,-20){2}
    \Vertex(9,-20){2}
    \Vertex(128,18){4}
    \Vertex(64,18){4}
    \Vertex(96,130){2}
    \Vertex(96,66){4}
    \Text(103,125)[lb]{$x_1$}
    \Text(-7,-22)[lb]{$x_2$}
    \Text(188,-22)[lb]{$x_3$}
  \end{picture}
\end{center}
where we used the black blob propagator notation introduced in subsection \ref{sec:phi phi s}. 

To avoid over-counting the contribution of the leading order  triangle diagram, it should be subtracted from the above graphs. The remainder contributes to $W_{s^3}\sim\mathcal{O}(1/N)$ which is the ultimate goal of our calculation.  However, to avoid clutter we do not carry out these subtractions explicitly. They are done by default in what follows.

Remarkably, when summing the Feynman diagrams with identical skeleton structure, the anomalous dimensions of the propagators behave additively at $1/N$ order. This tremendous simplification holds because $\gamma_{\phi,s}\sim 1/N$, and therefore one can linearize a Feynman graph with respect to the anomalous dimensions. 

For instance, each of the three dressed vertices of the leading order triangle diagram contributes a $1/N$ term associated with the second diagram on the r.h.s. of (\ref{Dressed vertex diagram}). Combining these terms with the above Feynman graph gives

\begin{center}
  \begin{picture}(162,162) (15,-31)
    \SetWidth{1.0}
    \SetColor{Black}
    \Line[](96,130)(96,66)
    \Line[](96,66)(64,18)
    \Line[](96,66)(128,18)
    \Line[](64,18)(128,18)
    \Line[](128,18)(182,-20)
    \Line[](64,18)(9,-20)
    \Vertex(182,-20){2}
    \Vertex(9,-20){2}
    \Vertex(128,18){4}
    \Vertex(64,18){4}
    \Vertex(96,130){2}
    \Vertex(96,66){4}
    \Text(82,21)[lb]{\scalebox{0.6}{$2(\Delta_\phi-\gamma_\phi)$}}
    \Text(46,43)[lb]{\scalebox{0.6}{$2(\Delta_\phi-\gamma_\phi)$}}
    \Text(116,43)[lb]{\scalebox{0.6}{$2(\Delta_\phi-\gamma_\phi)$}}
    \Text(100,95)[lb]{\scalebox{0.6}{$2\Delta_s$}}
    \Text(20,0)[lb]{\scalebox{0.6}{$2\Delta_s$}}
    \Text(158,0)[lb]{\scalebox{0.6}{$2\Delta_s$}}
    \Text(103,125)[lb]{$x_1$}
    \Text(103,65)[lb]{$x_6$}
    \Text(47,17)[lb]{$x_4$}
    \Text(135,17)[lb]{$x_5$}
    \Text(-7,-22)[lb]{$x_2$}
    \Text(188,-22)[lb]{$x_3$}
  \end{picture}
\end{center}
Since this relation between the diagrams is reliable up to $\mathcal{O}(1/N)$ order, we linearize over the $\gamma_\phi$ in the internal $\phi$-propagators and retain the next-to-leading corrections only. They are given by the sum of three diagrams which are identical up to a permutation of the external legs
\begin{center}
\raggedright
  \begin{picture}(162,162) (15,-31)
    \SetWidth{1.0}
    \SetColor{Black}
    \Line[](146,130)(114,18)
    \Line[](146,130)(178,18)
    \Line[](114,18)(178,18)
    \Line[](178,18)(232,-20)
    \Line[](114,18)(59,-20)
    \Vertex(232,-20){2}
    \Vertex(59,-20){2}
    \Vertex(178,18){4}
    \Vertex(114,18){4}
    \Vertex(146,130){2}
    \Text(123,21)[lb]{\scalebox{0.6}{$4\Delta_\phi-\Delta_s-2\gamma_\phi$}}
    \Text(116,83)[lb]{\scalebox{0.6}{$\Delta_s$}}
    \Text(166,83)[lb]{\scalebox{0.6}{$\Delta_s$}}
    \Text(70,0)[lb]{\scalebox{0.6}{$2\Delta_s$}}
    \Text(208,0)[lb]{\scalebox{0.6}{$2\Delta_s$}}
    \Text(153,125)[lb]{$x_1$}
    \Text(97,17)[lb]{$x_4$}
    \Text(185,17)[lb]{$x_5$}
    \Text(43,-22)[lb]{$x_2$}
    \Text(238,-22)[lb]{$x_3$}
    \Text(208,43)[lb]{\scalebox{1}{$+\quad$ two cyclic permutations $x_1\rightarrow x_2\rightarrow x_3\rightarrow x_1$.}} 
    \Text(45,43)[lb]{\scalebox{1}{$4C_{\phi\phi s}C_\phi C_s^2~\times$}} 
    \end{picture}
\end{center}
where we integrated over the unique vertex with the lines $\Delta_\phi$,
$\Delta_\phi$, $\Delta_s$ and used (\ref{Original phi phi s}). The factor of
$\left(-\frac{2}{\sqrt{N}}\right)^2NC_\phi C_s^{2}$
is associated with the Feynman rules for the vertices $x_{4,5}$, the closed $\phi$-loop, and the leading order amplitudes
of the $\phi$- and $s$- propagators.

Furthermore, there is a contribution related to the conformal triangle on the r.h.s. of (\ref{Dressed vertex diagram}). There are three such terms, since there are three vertices in the leading order correlation function
$\langle sss\rangle$,
\begin{center}
  \begin{picture}(162,162) (15,-31)
    \SetWidth{1.0}
    \SetColor{Black}
    \Line[](6,130)(-26,18)
    \Line[](6,130)(38,18)
    \Line[](-26,18)(38,18)
    \Line[](38,18)(92,-20)
    \Line[](-26,18)(-81,-20)
    \Vertex(92,-20){2}
    \Vertex(-81,-20){2}
    \Vertex(38,18){4}
    \Vertex(-26,18){4}
    \Vertex(6,130){2}
    \Text(-25,5)[lb]{\scalebox{0.6}{$4\Delta_\phi-\Delta_s+2\gamma_\phi-\gamma_s$}}
    \Text(-39,75)[lb]{\scalebox{0.6}{$\Delta_s+\gamma_s$}}
    \Text(27,75)[lb]{\scalebox{0.6}{$\Delta_s+\gamma_s$}}
    \Text(-70,0)[lb]{\scalebox{0.6}{$2\Delta_s$}}
    \Text(68,0)[lb]{\scalebox{0.6}{$2\Delta_s$}}
    \Text(13,125)[lb]{$x_1$}
    \Text(-43,17)[lb]{$x_4$}
    \Text(45,17)[lb]{$x_5$}
    \Text(-97,-22)[lb]{$x_2$}
    \Text(98,-22)[lb]{$x_3$}
    \Text(78,43)[lb]{\scalebox{1}{$+\quad$ two cyclic permutations $x_1\rightarrow x_2\rightarrow x_3\rightarrow x_1$.}} 
    \Text(-130,43)[lb]{\scalebox{1}{$4C_{\phi\phi s}(1+\delta V)C_\phi C_s^2~  \times$}} 
  \end{picture}
\end{center}
As before only $\mathcal{O}(1/N)$ terms are eventually retained, and therefore the last two diagrams can be combined by simply adding the anomalous dimensions of the corresponding propagators
\begin{center}
  \begin{picture}(415,162) (15,-31)
    \SetWidth{1.0}
    \SetColor{Black}
    \Line[](96,130)(64,18)
    \Line[](96,130)(128,18)
    \Line[](64,18)(128,18)
    \Line[](128,18)(182,-20)
    \Line[](64,18)(9,-20)
    \Vertex(182,-20){2}
    \Vertex(9,-20){2}
    \Vertex(128,18){4}
    \Vertex(64,18){4}
    \Vertex(96,130){2}
    \Text(78,21)[lb]{\scalebox{0.6}{$4\Delta_\phi-\Delta_s-\gamma_s$}}
    \Text(53,68)[lb]{\scalebox{0.6}{$\Delta_s+\gamma_s$}}
    \Text(116,68)[lb]{\scalebox{0.6}{$\Delta_s+\gamma_s$}}
    \Text(20,0)[lb]{\scalebox{0.6}{$2\Delta_s$}}
    \Text(158,0)[lb]{\scalebox{0.6}{$2\Delta_s$}}
    \Text(103,125)[lb]{$x_1$}
    \Text(47,17)[lb]{$x_4$}
    \Text(135,17)[lb]{$x_5$}
    \Text(-7,-22)[lb]{$x_2$}
    \Text(188,-22)[lb]{$x_3$}
    \Text(200,26)[lb]{\scalebox{1}{
    $=
     \frac{4 \, C_{\phi\phi s}\,C_\phi C_s^2\,
     U\left(\Delta_s,\frac{\Delta_s+\gamma_s}{2},d-\frac{3\Delta_s+\gamma_s}{2}\right)
     U\left(\Delta_\phi+\frac{\gamma_s}{2},\Delta_\phi-\frac{\gamma_s}{2},\Delta_s\right)}
     {\left(|x_{13}||x_{12}|\right)^{\Delta_s+\gamma_s}
|x_{23}|^{\Delta_s-\gamma_s}}$}}
 \Text(140,32)[lb]{\scalebox{1}{+  two perm.}} 
 \Text(285,12)[lb]{\scalebox{1}{$\times (1+\delta V)$ ~ +  ~ two permutations}} 
    \Text(-10,33)[lb]{\scalebox{0.7}{$4C_{\phi\phi s}(1+\delta V)C_\phi C_s^2 ~ \times$}} 
  \end{picture}
\end{center}
where the integrals over $x_{4,5}$ were carried out using the uniqueness relation. Note that only terms up to $\mathcal{O}(1/N)$ order are reliable, because we added the anomalous dimensions to combine the diagrams. Using (\ref{relation 1}), (\ref{Cs3 and Cphiphi s relation}), the contribution to $W_{s^3}$ takes the form
\begin{eqnarray}
\label{hat f expression}
3\delta V + \hat f &=& 3\left(\delta V + 
\frac{U\left(\Delta_s,\frac{\Delta_s+\gamma_s}{2},d-\frac{3\Delta_s+\gamma_s}{2}\right)
U(\Delta_\phi+\frac{\gamma_s}{2},\Delta_\phi-\frac{\gamma_s}{2},\Delta_s)}
{U\left(\Delta_s,\frac{\Delta_s}{2},d-\frac{3\Delta_s}{2}\right)
U(\Delta_\phi,\Delta_\phi,\Delta_s)} -1 \right)
\nonumber \\
&=&3\delta V + \frac{6 \sin \left(\frac{\pi  d}{2}\right) \Gamma (d) \left(-\frac{2}{d-4}+\pi  \cot \left(\frac{\pi  d}{2}\right)
+H_{d-4} \right)}{ N \pi  \Gamma \left(\frac{d}{2}-1\right) \Gamma \left(\frac{d}{2}+1\right)} 
+ \mathcal{O}(1/N^2)\,,
\end{eqnarray}
where $H_n$ is the $n$\textit{th} harmonic number. Expanding around $d=6$, yields
\begin{equation}
\label{f around 6d}
\hat f = -{120\over N} +{\cal O}(d-6)\,.
\end{equation}

Another contribution to $W_{s^3}$ arises from the following diagram
\begin{center}
  \begin{picture}(162,162) (15,-31)
    \SetWidth{1.0}
    \SetColor{Black}
    \Line[](96,130)(96,98)
    \Line[](37,0)(155,0)
    \Line[](155,0)(182,-20)
    \Line[](37,0)(9,-20)
    \Line[](37,0)(96,98)
    \Line[](155,0)(96,98)
    \Vertex(182,-20){2}
    \Vertex(9,-20){2}
    \Vertex(155,0){4}
    \Vertex(37,0){4}
    \Vertex(96,130){2}
    \Vertex(96,98){4}
    \Line[](78,68)(114,68)
    \Vertex(78,68){4}
    \Vertex(114,68){4}
    \Line[](75,0)(56,32)
    \Line[](117,0)(136,32)
    \Vertex(75,0){4}
    \Vertex(56,32){4}
    \Vertex(117,0){4}
    \Vertex(136,32){4}
    \Text(100,112)[lb]{\scalebox{0.6}{$2\Delta_s$}}
    \Text(8,-10)[lb]{\scalebox{0.6}{$2\Delta_s$}}
    \Text(171,-10)[lb]{\scalebox{0.6}{$2\Delta_s$}}
    \Text(103,125)[lb]{$x_1$}
    \Text(103,98)[lb]{$x_7$}
    \Text(20,0)[lb]{$x_8$}
    \Text(163,0)[lb]{$x_9$}
    \Text(-7,-22)[lb]{$x_2$}
    \Text(188,-22)[lb]{$x_3$}
    \Text(61,68)[lb]{$x_4$}
    \Text(73,-15)[lb]{$x_5$}
    \Text(143,32)[lb]{$x_6$}
    \Text(51,-10)[lb]{\scalebox{0.6}{$2\Delta_\phi$}}
    \Text(91,-10)[lb]{\scalebox{0.6}{$2\Delta_s$}}
    \Text(131,-10)[lb]{\scalebox{0.6}{$2\Delta_\phi$}}
    \Text(30,15)[lb]{\scalebox{0.6}{$2\Delta_\phi$}}
    \Text(70,15)[lb]{\scalebox{0.6}{$2\Delta_\phi$}}
    \Text(110,15)[lb]{\scalebox{0.6}{$2\Delta_\phi$}}
    \Text(150,15)[lb]{\scalebox{0.6}{$2\Delta_\phi$}}
    \Text(52,50)[lb]{\scalebox{0.6}{$2\Delta_s$}}
    \Text(130,50)[lb]{\scalebox{0.6}{$2\Delta_s$}}
    \Text(90,58)[lb]{\scalebox{0.6}{$2\Delta_\phi$}}
    \Text(71,82)[lb]{\scalebox{0.6}{$2\Delta_\phi$}}
    \Text(109,82)[lb]{\scalebox{0.6}{$2\Delta_\phi$}}
  \end{picture}
\end{center}
Integrating over $x_{4-9}$ through the use of uniqueness relation (\ref{uniqueness}),
and normalizing the external $s$ legs according to  (\ref{field normalizations}),
we obtain an additional contribution to $W_{s^3}$
of the form $w_3/\tilde C_{s^3}$ with
\begin{align}
\label{w3}
w_3&=\frac{1}{C_s^{3/2}}\,\left(-\frac{2}{\sqrt{N}}\right)^9\,N^3\,
C_\phi^9 C_s^6\,U\left(\Delta_\phi,\Delta_\phi,\Delta_s\right)^3
U\left(\frac{\Delta_s}{2}, \Delta_s, d-\frac{3\Delta_s}{2}\right)^3\,\hat w_3\notag\\
&=-\frac{64(d-3)^3}{N^{3/2}}\,C_\phi^3\,C_s^\frac{3}{2}\,\hat w_3\,,
\end{align}
where in the last line we used (\ref{relation 1}), and
$\hat w_3$ is defined by the diagram\footnote{From now on we
explicitly use the values (\ref{engineering scaling}) for the scaling dimensions $\Delta_{\phi, s}$. To get $\hat w_3$ from this diagram one needs to integrate over the three internal points.
We already accounted for the amplitudes $C_{\phi, s}$ of the propagators and factors of $-2/\sqrt{N}$ coming from the interaction vertices.}
\begin{center}
  \begin{picture}(165,150) (65,-5)
    \SetWidth{1.0}
    \SetColor{Black}
    \Line[](32,6)(192,6)
    \Line[](192,6)(112,134)
    \Line[](112,134)(32,6)
    \Line[](112,6)(72,74)
    \Line[](72,70)(152,70)
    \Line[](152,70)(112,6)
    \Vertex(72,70){4}
    \Vertex(152,70){4}
    \Vertex(112,134){2}
    \Vertex(32,6){2}
    \Vertex(112,6){4}
    \Vertex(192,6){2}
    \Text(120,135)[lb]{$x_1$}
    \Text(17,-5)[lb]{$x_2$}
    \Text(198,-5)[lb]{$x_3$}
    \Text(70,103)[lb]{\scalebox{0.6}{$6-d$}}
    \Text(65,-5)[lb]{\scalebox{0.6}{$6-d$}}
    \Text(183,35)[lb]{\scalebox{0.6}{$6-d$}}
    \Text(105,76)[lb]{\scalebox{0.6}{$d-2$}}
    \Text(72,35)[lb]{\scalebox{0.6}{$d-2$}}
    \Text(137,35)[lb]{\scalebox{0.6}{$d-2$}}
    \Text(145,-5)[lb]{\scalebox{0.6}{$d-2$}}
    \Text(28,35)[lb]{\scalebox{0.6}{$d-2$}}
    \Text(140,103)[lb]{\scalebox{0.6}{$d-2$}}
    \Text(210,62)[lb]{$=\frac{\hat w_3}{|x_{12}|^{2}|x_{13}|^{2}|x_{23}|^{2}}$}
  \end{picture}
\end{center}
Integrating both sides of this diagrammatic equation w.r.t. $x_1$ we obtain
\begin{equation}
\label{hat w3 in terms of v3}
\hat w_3=\frac{U\left(3-\frac{d}{2},\frac{d}{2}-1,d-2\right)}{U(1,1,d-2)}\,v_3\,,
\end{equation}
where $v_3$ is determined by the diagram
\begin{center}
  \begin{picture}(165,90) (75,-5)
    \SetWidth{1.0}
    \SetColor{Black}
    \Line[](32,6)(192,6)
    \Line[](192,6)(152,70)
    \Line[](72,70)(32,6)
    \Line[](112,6)(72,74)
    \Line[](72,70)(152,70)
    \Line[](152,70)(112,6)
    \Vertex(72,70){4}
    \Vertex(152,70){4}
    \Vertex(32,6){2}
    \Vertex(112,6){4}
    \Vertex(192,6){2}
    \Text(17,-5)[lb]{$x_2$}
    \Text(198,-5)[lb]{$x_3$}
    \Text(65,-5)[lb]{\scalebox{0.6}{$6-d$}}
    \Text(183,35)[lb]{\scalebox{0.6}{$6-d$}}
    \Text(110,76)[lb]{\scalebox{0.6}{$2$}}
    \Text(72,35)[lb]{\scalebox{0.6}{$d-2$}}
    \Text(137,35)[lb]{\scalebox{0.6}{$d-2$}}
    \Text(145,-5)[lb]{\scalebox{0.6}{$d-2$}}
    \Text(28,35)[lb]{\scalebox{0.6}{$d-2$}}
    \Text(220,30)[lb]{$=\frac{v_3}{|x_{23}|^{6-d}}$}
  \end{picture}
\end{center}
Attaching propagator lines with powers $2d-4$ to $x_{2,3}$, and integrating
both sides of the obtained equation w.r.t. $x_{2,3}$, yields
\begin{equation}
\label{v3 in terms of tilde v3}
v_3 = \frac{U\left(3-\frac{d}{2},\frac{d}{2}-1,d-2\right)}{U(1,1,d-2)}\,\tilde v_3(0)\,,
\end{equation}
where $\tilde v_3(\delta)$ is defined by the diagram
\begin{center}
  \begin{picture}(165,90) (75,-5)
    \SetWidth{1.0}
    \SetColor{Black}
    \Line[](32,6)(192,6)
    \Line[](192,6)(152,70)
    \Line[](72,70)(32,6)
    \Line[](112,6)(72,74)
    \Line[](72,70)(152,70)
    \Line[](152,70)(112,6)
    \Vertex(72,70){4}
    \Vertex(152,70){4}
    \Vertex(32,6){2}
    \Vertex(112,6){4}
    \Vertex(192,6){2}
    \Text(17,-5)[lb]{$x_2$}
    \Text(198,-5)[lb]{$x_3$}
    \Text(68,-5)[lb]{\scalebox{0.6}{$2$}}
    \Text(183,35)[lb]{\scalebox{0.6}{$2-\delta$}}
    \Text(110,76)[lb]{\scalebox{0.6}{$2$}}
    \Text(82,35)[lb]{\scalebox{0.6}{$2$}}
    \Text(137,35)[lb]{\scalebox{0.6}{$2$}}
    \Text(145,-5)[lb]{\scalebox{0.6}{$2d-6$}}
    \Text(10,35)[lb]{\scalebox{0.6}{$2d-6+2\delta$}}
    \Text(220,30)[lb]{$=\frac{\tilde v_3(\delta )}{|x_{23}|^{d-2+\delta}}$}
  \end{picture}
\end{center}
Here we have introduced an auxiliary regulator, $\delta$, in order to apply the integration by parts
relation (the diagram itself is finite in $\delta\rightarrow 0$ limit). We relegate the details to 
Appendix \ref{appendix: calculation of v3}.
Combining (\ref{phi phi s leading}), (\ref{Cs3 and Cphiphi s relation}), (\ref{w3}), (\ref{hat w3 in terms of v3}),
(\ref{v3 in terms of tilde v3}) gives the result
\begin{equation}
\label{w3 over Cs3}
\frac{w_3}{\tilde C_{s^3}} = \frac{16(d-3)^2}{N}\,C_\phi^2 C_s
\, \frac{U\left(3-\frac{d}{2},\frac{d}{2}-1,d-2\right)^2}{U(1,1,d-2)^2\,
U\left(\frac{d-2}{2},\frac{d-2}{2},2\right)}\,\tilde  v_3(0)\,,
\end{equation}
where $\tilde  v_3(0)$ is given by (\ref{tilde v3 answer}).

\begin{figure}[htb]
\begin{center}
\includegraphics[width=300pt]{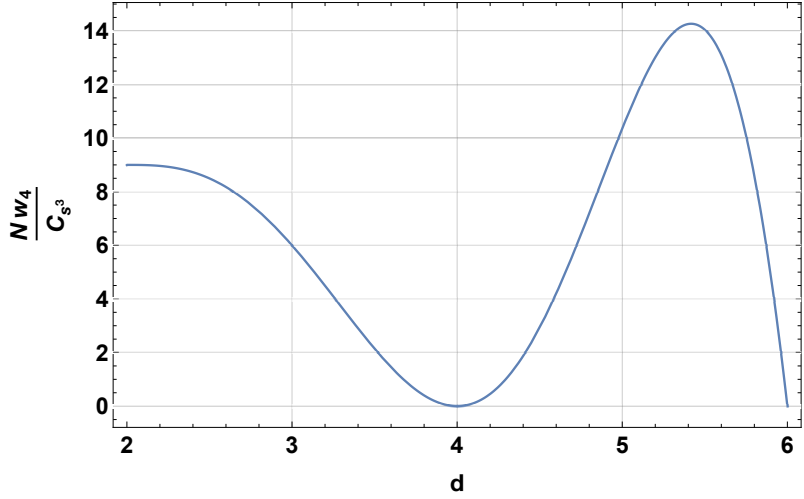}
\end{center}
\caption{$Nw_4/\tilde C_{s^3}$ as a function of space-time dimension $2<d<6$.}
\label{fig:w4}
\end{figure}

Next, we evaluate the following contribution to $W_{s^3}$ represented by the 3-loop `bellows'\footnote{The name is based on the visual resemblance of the three-dimensional shape of this diagram to a bellows, see, \textit{e.g.}, \href{https://en.wikipedia.org/wiki/Bellows}{\underline{https://en.wikipedia.org/wiki/Bellows}}} diagram\footnote{We are grateful to anonymous referee at PRD who pointed out this diagram to us.}
\begin{center}
  \begin{picture}(324,136) (47,-39)
    \SetWidth{1.0}
    \SetColor{Black}
    \Line[](28,22)(94,22)
    \Line[](172,76)(172,-32)
    \Line[](172,22)(208,22)
    \Line[](250,22)(172,76)
    \Line[](250,22)(172,-32)
    \Line[](250,22)(316,22)
    \Line[](94,22)(172,76)
    \Line[](94,22)(172,-32)
    \Line[](136,52)(136,-8)
    \Vertex(136,52){4}
    \Vertex(136,-8){4}
    \Vertex(172,76){4}
    \Vertex(172,-32){4}
    \Vertex(172,22){4}
    \Vertex(94,22){4}
    \Vertex(250,22){4}
    \Vertex(30,22){2}
    \Vertex(208,22){2}
    \Vertex(316,22){2}
    \Text(25,28)[lb]{$x_1$}
    \Text(312,28)[lb]{$x_2$}
    \Text(206,28)[lb]{$x_3$}
    \Text(85,28)[lb]{$x_4$}
    \Text(130,60)[lb]{$x_5$}
    \Text(130,-23)[lb]{$x_6$}
    \Text(178,28)[lb]{$x_7$}
    \Text(253,28)[lb]{$x_8$}
    \Text(58,28)[lb]{\scalebox{0.6}{$2\Delta_s$}}%
    \Text(106,42)[lb]{\scalebox{0.6}{$2\Delta_\phi$}}
    \Text(106,-2)[lb]{\scalebox{0.6}{$2\Delta_\phi$}}
    \Text(142,22)[lb]{\scalebox{0.6}{$2\Delta_\phi$}}
    \Text(147,70)[lb]{\scalebox{0.6}{$2\Delta_s$}}
    \Text(147,-29)[lb]{\scalebox{0.6}{$2\Delta_s$}}
    \Text(192,28)[lb]{\scalebox{0.6}{$2\Delta_s$}}%
    \Text(178,45)[lb]{\scalebox{0.6}{$2\Delta_\phi$}}
    \Text(178,-8)[lb]{\scalebox{0.6}{$2\Delta_\phi$}}
    \Text(214,51)[lb]{\scalebox{0.6}{$2\Delta_\phi$}}
    \Text(214,-14)[lb]{\scalebox{0.6}{$2\Delta_\phi$}}
    \Text(280,28)[lb]{\scalebox{0.6}{$2\Delta_s$}}%
    \Text(332,12)[lb]{$=\frac{w_4}{(|x_{12}||x_{13}||x_{23}|)^{\Delta_s}}$}
  \end{picture}
\end{center}
Normalizing the external $s$ legs in accord with (\ref{field normalizations}),
and using the uniqueness relation (\ref{uniqueness}) to 
integrate over the five points $x_{4-8}$, yields\footnote{The symmetry factor of this diagram is $1/2$, whereas its multiplicity is $3$
due to the possibility of having the $\phi$-triangle attached to each of the points $x_{1-3}$. Recall also that the contribution to $W_{s^3}$ is obtained by dividing the value of $w_4$ by $\tilde C_{s^3}$, see (\ref{sss general}).}
\begin{equation}
\label{w3}
W_{s^3}\supset\frac{w_4}{\tilde C_{s^3}} = \frac{3}{2} \Big( -\frac{2}{\sqrt{N}} \Big)^4 \,N\,
C_\phi^4 C_s^2\,U(\Delta_\phi,\Delta_\phi,\Delta_s)^2\, v_4\,,
\end{equation}
where $v_4$ is defined by the followihg diagrammatic equation
(the multiplicative factors of $-2/\sqrt{N}$ and $C_{\phi, s}$ in the vertices and propagators of this diagram should be stripped off, because
we already took them into account)
\begin{center}
  \begin{picture}(374,136) (47,-39)
    \SetWidth{1.0}
    \SetColor{Black}
    \Line[](192,76)(192,-32)
    \Line[](270,22)(192,76)
    \Line[](270,22)(192,-32)
    \Line[](114,22)(192,76)
    \Line[](114,22)(192,-32)
     \Line[](192,76)(228,22)
    \Line[](192,-32)(228,22)
    \Vertex(192,76){4}
    \Vertex(192,-32){4}
    \Vertex(114,22){2}
    \Vertex(270,22){2}
    \Vertex(228,22){2}
    \Text(226,28)[lb]{$x_3$}
    \Text(105,28)[lb]{$x_1$}
    \Text(273,28)[lb]{$x_2$}
    \Text(300,12)[lb]{$=\frac{v_4}{(|x_{12}||x_{13}||x_{23}|)^{\Delta_s}}$}
    \Text(140,52)[lb]{\scalebox{0.6}{$\Delta_s$}}
    \Text(140,-12)[lb]{\scalebox{0.6}{$\Delta_s$}}
    \Text(162,22)[lb]{\scalebox{0.6}{$2d-3\Delta_s$}}
    \Text(205,40)[lb]{\scalebox{0.6}{$\Delta_s$}}
    \Text(205,3)[lb]{\scalebox{0.6}{$\Delta_s$}}
    \Text(234,51)[lb]{\scalebox{0.6}{$\Delta_s$}}
    \Text(234,-14)[lb]{\scalebox{0.6}{$\Delta_s$}}
  \end{picture}
\end{center}
Integrating both sides of this equation w.r.t. $x_3$, we obtain
(here we explicitly use the values (\ref{engineering scaling}) for the scaling dimensions $\Delta_{\phi, s}$)
\begin{center}
  \begin{picture}(374,136) (47,-39)
    \SetWidth{1.0}
    \SetColor{Black}
    \Line[](192,76)(192,-32)
    \Line[](270,22)(192,76)
    \Line[](270,22)(192,-32)
    \Line[](114,22)(192,76)
    \Line[](114,22)(192,-32)
    \Vertex(192,76){4}
    \Vertex(192,-32){4}
    \Vertex(114,22){2}
    \Vertex(270,22){2}
    \Text(105,28)[lb]{$0$}
    \Text(273,28)[lb]{$x$}
    \Text(300,12)[lb]{$=\frac{v_4}{|x|^{6-d}}$}
    \Text(140,52)[lb]{\scalebox{0.6}{$2$}}
    \Text(140,-12)[lb]{\scalebox{0.6}{$2$}}
    \Text(168,22)[lb]{\scalebox{0.6}{$d-2$}}
    \Text(234,51)[lb]{\scalebox{0.6}{$2$}}
    \Text(234,-14)[lb]{\scalebox{0.6}{$2$}}
  \end{picture}
\end{center}
This is the so-called self-energy diagram. It can be reduced to the known
$\textrm{ChT}(\alpha,\beta)$ graph, given by eq. (16) in \cite{Vasiliev:1981dg}.\footnote{See Appendix
\ref{appendix: calculation of v3} for details regarding the self-energy diagram, {\it e.g.,}
(\ref{ChT expression}) for an explicit form of $\textrm{ChT}(\alpha,\beta)$.}
To this end we perform an inversion transformation on the external point $x$
and on both of the integrated vertices. This gives $v_4 = \textrm{ChT}(1,1)$. Combining
all together, we arrive at
\begin{align}
\label{w4 result}
\frac{w_4}{\tilde C_{s^3}}=\gamma_\phi\,\frac{3 d (d-2)  \left(\pi ^2-6 \psi ^{(1)}\left(\frac{d}{2}-1\right)\right)}{4 (d-4)}\,,
\end{align}
where $\psi^{(1)}$ is the first derivative of the digamma function. In particular, 
\begin{align}
\label{w4 in d4}
\frac{w_4}{\tilde C_{s^3}} &=-\frac{9}{2}\,\frac{1}{N}\, (d-4)^2 \psi ^{(2)}(1)+{\cal O}\left((d-4)^3\right)\,,\qquad d\rightarrow 4\,,\\
\label{w4 in d6}
\frac{w_4}{\tilde C_{s^3}} &= -54\, \frac{1}{N}\,(d-6) + {\cal O}((d-6)^{2})\,,\qquad d\rightarrow 6\,.
\end{align}
The plot of (\ref{w4 result}) for the range of values $2<d<6$ is shown in figure \ref{fig:w4}.

To account for the contribution of the $\mathcal{O}(1/N)$ corrections, $A_{\phi, s}$, to the amplitudes 
of the dressed $\phi$- and $s$-propagators, we simply add to $W_{s^3}$ a term $3A_s+3A_\phi$.
Furthermore, there is an additional term $-\frac{3}{2}A_s$, associated with normalization (\ref{field normalizations}) of the three external $s$ legs. As a result, $W_{s^3}$ takes the form
\begin{equation}
\label{Ws3 total}
W_{s^3} =3\,\delta V + \hat f+\frac{3}{2}A_s+3A_\phi  + \frac{w_3+w_4}{\tilde C_{s^3}}\,.
\end{equation}
Substituting (\ref{general W phi phi s}), yields
\begin{equation}
\label{Ws3 total in terms of Wpps}
W_{s^3} = 3W_{\phi\phi s} +\hat f+ \frac{w_3+w_4}{\tilde C_{s^3}}\,.
\end{equation}
Here $w_3/\tilde C_{s^3}$ is given by (\ref{w3 over Cs3}), which we re-write as
\begin{equation}
\label{w3 over Cs3 in terms of b3 and tilde v3}
\frac{w_3}{\tilde C_{s^3}} = b_3\,\tilde v_3(0)\,,
\end{equation}
where $\tilde v_3(0)$ is calculated in Appendix \ref{appendix: calculation of v3}, and
\begin{equation}
\label{b3 final answer}
b_3 = \frac{16(d-3)^2}{N}\,C_\phi^2 C_s
\, \frac{U\left(3-\frac{d}{2},\frac{d}{2}-1,d-2\right)^2}{U(1,1,d-2)^2\, ~
U\left(\frac{d-2}{2},\frac{d-2}{2},2\right)}\,.
\end{equation}
In particular, it simplifies in the vicinity of $d=4,\, 6$
\begin{align}
\label{b3 in d4}
b_3 &= \frac{1}{N}\,\frac{(d-4)^3}{\pi^6} + {\cal O}((d-4)^4)\,,\qquad d\rightarrow 4\,,\\
\label{b3 in d6}
b_3 &= - \frac{1}{N} \,\frac{216(d-6)^3}{\pi^9} + {\cal O}((d-6)^4)\,,\qquad d\rightarrow 6\,.
\end{align}
Moreover, from (\ref{tilde v3 answer}) we obtain
\begin{align}
\label{tilde v3 in d4}
\tilde v_3(0) &= 20\pi^6\zeta(5) + {\cal O}(d-4)\,,\qquad d\rightarrow 4\,,\\
\label{tilde v3 in d6}
\tilde v_3(0) &= - \frac{\pi^9}{(d-6)^3} + {\cal O}((d-6)^{-2})\,,\qquad d\rightarrow 6\,.
\end{align}

\begin{figure}[htb]
\begin{center}
\includegraphics[width=300pt]{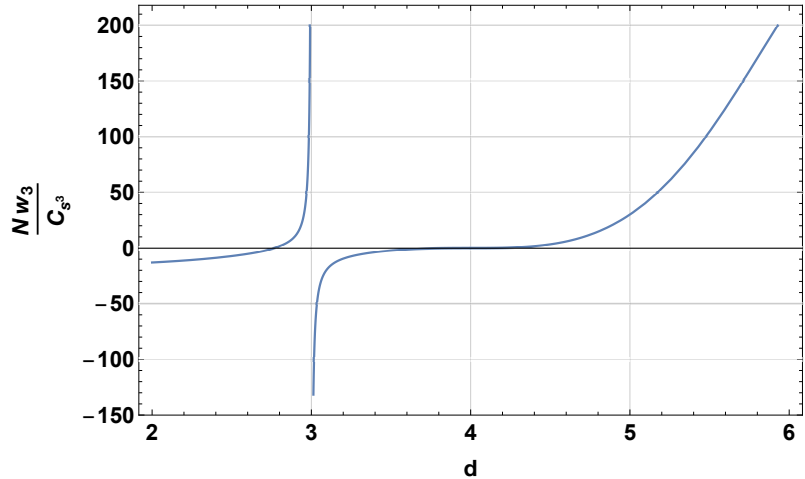}
\end{center}
\caption{$Nw_3/\tilde C_{s^3}$ as a function of space-time dimension $2<d<6$. Our choice of normalization results in a pole, because $\tilde C_{s^3}$ vanishes in $d=3$.}
\label{fig:w3}
\end{figure}

The full expression for $w_3/\tilde C_{s^3}$ is displayed in
(\ref{w3 over C3 full result}) below, where $\psi^{(n)}(x)$ is \textit{n}th derivative of the digamma function $\psi^{(0)}(x)=\Gamma'(x)/\Gamma(x)$,
and $\gamma$ is the Euler constant. The plot of (\ref{w3 over C3 full result}) for the range $2\leq d\leq 6$ is shown in figure~\ref{fig:w3}. There is an apparent singularity in $d=3$, which is an artifact of our normalization. Note that $\tilde C_{s^3}$  has a simple zero in $d=3$.

The theory becomes free at the fixed point in the limit $d\to 4$, therefore $W_{s^3}(d\rightarrow 4)$ must vanish.\footnote{We would like to thank Simone Giombi, Igor Klebanov and Gregory Tarnopolsky for discussing with us the $d=4$ case. Their insight helped us to improve our previous version of the related calculation.}
Indeed, based on our (\ref{general W phi phi s}), $W_{\phi\phi s}(d\rightarrow 4) = 0$, in agreement with \cite{Petkou:1995vu,Petkou:1994ad}.
From (\ref{Ws3 total in terms of Wpps}) it then remains to show that
$\frac{w_3+w_4}{\tilde C_{s^3}}(d\rightarrow 4) = 0$, which is indeed the case
(in fact $w_{3,4}/\tilde C_{s^3}$ contributions vanish individually), as can be seen from
(\ref{w4 in d4}), (\ref{w3 over Cs3 in terms of b3 and tilde v3}),
(\ref{b3 in d4}), (\ref{tilde v3 in d4}). Using (\ref{hat f expression}) one can also verify 
that $\hat f(d = 4) = 0$.

Moreover, it follows from (\ref{w3 over Cs3 in terms of b3 and tilde v3}),
(\ref{b3 in d6}), (\ref{tilde v3 in d6}) that in the vicinity of $d=6$, we have
\begin{equation}
\label{w3 over Cs3 in d6}
\frac{w_3}{\tilde C_{s^3}} = \frac{216}{N} + {\cal O}(d-6)\,.
\end{equation}
From (\ref{d6 W phi phi s}), (\ref{f around 6d}),
 (\ref{w4 in d6}), (\ref{Ws3 total in terms of Wpps}) and (\ref{w3 over Cs3 in d6})
we then obtain in $d=6-\epsilon$,
\begin{equation}
\label{Ws3 final result}
W_{s^3} = \frac{162}{N} + {\cal O}(\epsilon)\,.
\end{equation}
The same value for $W_{s^3}$ was obtained in  \cite{Fei:2014yja,Fei:2014xta}
for the critical cubic model. This match between the OPE coefficients provides an additional non-trivial evidence for the equivalence between the models. In fact, it suggests existence of a wide class of universal relations between the $O(N)$ CFTs in general $d$, at least in the $1/N$ expansion. Some of these relations
have been established in \cite{Petkou:1995vu,Petkou:1994ad} using the bootstrap method and
without considering a particular
Lagrangian or space-time dimension.
Our findings suggest that the bootstrap approach to the $O(N)$ CFTs, initiated in
\cite{Petkou:1995vu,Petkou:1994ad}, might not be exhausted.\footnote{Recent advancements in 
bootstrap calculations in the $O(N)$ vector models have been reported in \cite{Alday:2019clp}.}
Presumably it can be extended to unravel a larger class of universal relations, which hold regardless of the specific structure of the $O(N)$ invariant  Hamiltonian at the fixed point. In fact, these relations could be valid
in general $d$, and apply to all $O(N)$ symmetric CFTs with the same number of degrees of freedom,
rather than just to the critical $\phi^4$
vector model or cubic model of \cite{Fei:2014yja,Fei:2014xta}.
A non-trivial match (\ref{Ws3 final result}) of the OPE data to $1/N$ order in the case of apparently distinct models provides a partial evidence to this statement.

\begin{figure}[htb]
\begin{center}
\includegraphics[width=300pt]{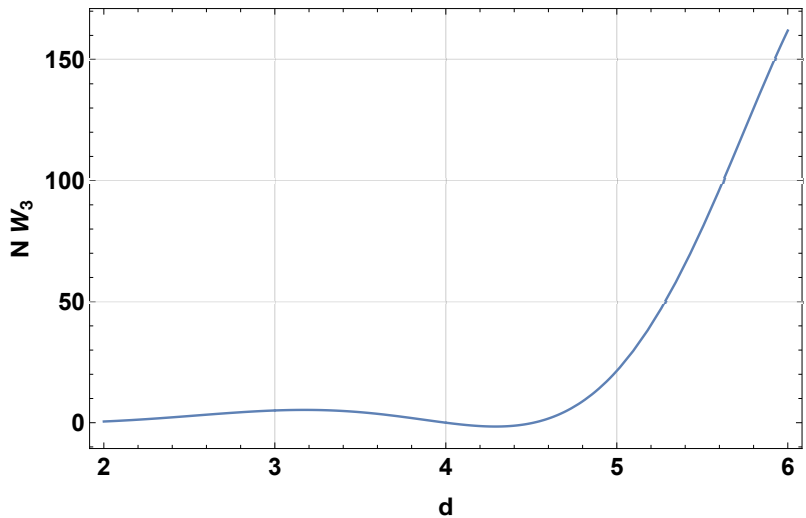}
\end{center}
\caption{The total $1/N$ correction $W_{s^3}$ 
to $\langle sss\rangle$ as a function of space-time dimension $2<d<6$. 
Some of the values are $NW_{s^3}(d=2)=1/2$,
$NW_{s^3}(d=4)=0$, $NW_{s^3}(d=6)=162$.}
\label{fig:W3Tot}
\end{figure}

We plot the full result (\ref{Ws3 total in terms of Wpps}) for $W_{s^3}$ in figure \ref{fig:W3Tot}.
It should be noticed that the lack of the refined result for the $\langle sss\rangle$ correlator
has been particularly emphatic since the work of \cite{Petkou:1995vu,Petkou:1994ad},
where the next-to-leading order value for the $\langle \phi\phi s\rangle$
three-point function was established. The new result
(\ref{Ws3 total in terms of Wpps}) for the $\langle ss s\rangle$
allowed us to subject the hypothesis of \cite{Fei:2014yja,Fei:2014xta} to a non-trivial test.

Finally, we would like to stress that while the results for the multi-loop diagrams, {\it e.g.,} 
the 4-loop triangle diagram (\ref{w3 over C3 full result}), played a crucial role in the direct
diagrammatic calculation of the $\langle sss\rangle$ three-point function
at the next-to-leading order in the $1/N$ expansion, they have independent value, because diagrams of this type are ubiquitous in perturbative CFTs. Two additional useful relations obtained in this section are the 3-loop trapezoid diagram $v_3$, or equivalently $\tilde v_3$ (see also Appendix \ref{appendix: calculation of v3}) and the 3-loop bellow diagram (\ref{w4 result}). To best of our knowledge the analytic expressions for these diagrams do not appear in the literature.

For convenience, we  recapitulate the full answer for $W_{s^3}$,
determined by (\ref{A phi 1over N general}), 
(\ref{A s 1over N general}),
(\ref{general W phi phi s}),
(\ref{delta V in general d}),
(\ref{hat f expression}),
(\ref{w4 result}),
(\ref{Ws3 total in terms of Wpps}) and
(\ref{w3 over Cs3 in terms of b3 and tilde v3})
\begin{align}
W_{s^3} &= 3W_{\phi\phi s} +\hat f+ \frac{w_3+w_4}{\tilde C_{s^3}}\,,\\
W_{\phi\phi s} &= \gamma_\phi\,\left(\frac{d(d-3)+4}{4-d}
\Big( H_{d-3}+\pi  \cot \big(\frac{\pi  d}{2}\big)\Big)
+\frac{16}{(d-4)^2}+\frac{6}{d-4}+\frac{2}{d-2}-2 d+3\right)\,,\\
%
%
\hat f&=\gamma_\phi\,\frac{6(d-1)(d-2)}{d-4}\,\left(H_{d-4}-\frac{2}{d-4}+\pi  \cot \big(\frac{\pi  d}{2}\big) \right)\,,\\
\frac{w_3}{\tilde C_{s^3}} &=\gamma_\phi {2d(d{-}2)(d{-}3)\over (d{-}4)^2} 
\left( 6 \psi ^{(1)}\Big(\frac{d}{2}{-}1\Big){-}\psi ^{(1)}(d{-}3) {-}{\pi^2\over 6}
{-} H_{d-4}\Big(H_{d{-}4}{+}2\pi\cot\big({\pi d\over 2}\big) \Big)
\right) 
\label{w3 over C3 full result}\,,\\
\frac{w_4}{\tilde C_{s^3}}&=\gamma_\phi\,\frac{3 d (d-2)  \left(\pi ^2-6 \psi ^{(1)}\left(\frac{d}{2}-1\right)\right)}{4 (d-4)}\,,
\end{align}
where $\psi^{(n)}$ is the $n$\textit{th} derivative of the digamma function
$\psi^{(0)}(x) = \Gamma'(x)/\Gamma(x)$,
$H_n$ is the $n$\textit{th} harmonic number,
and $\gamma_\phi\sim 1/N$ is given by (\ref{gamma phi 1over N general}).

We conclude this section by discussing a potential application of our results for the three-point function
$\langle sss\rangle$ in the context of $3d/4d$ critical vector model/higher spin theory correspondence
\cite{Klebanov:2002ja}.
We find that in $d=3$ the OPE coefficient vanishes up to the next-to-leading order in $1/N$, \footnote{Most
accurate available numerical values of the total $\langle sss\rangle$ OPE coefficient
$\lambda_{sss}$ in $d=3$ were presented in \cite{Kos:2016ysd,Chester:2019ifh}
\begin{equation}
\lambda_{sss}\Bigg|_{O(2)} = 0.830914(32)\,,\qquad
\lambda_{sss}\Bigg|_{O(3)}= 0.499(12)\,.
\end{equation}}
\begin{equation}
\label{Ws3 in d=3}
W_{s^3}(d=3) = 0+ {\cal O}(1/N^2)\,.
\end{equation}
It happens because of a rather non-trivial cancellation between the simple poles of two apparently unrelated terms in the expression for $W_{s^3}$
\begin{align}
N\, \hat f &= \frac{16}{\pi^2}\,\frac{1}{d-3}+{\cal O}((d-3)^0)\,,\\
N\, \frac{w_3}{\tilde C_{s^3}}&= - \frac{16}{\pi^2}\,\frac{1}{d-3}+{\cal O}((d-3)^0)\,.
\end{align}

A large body of the literature, starting from the work of \cite{Klebanov:2002ja}
was dedicated to studying the holographic
correspondence between the critical $O(N)$ vector model in $d=3$ and the type-A Vasiliev higher-spin theory
in $AdS_4$.
While in this paper we do not do justice to properly reviewing related literature, we would like
to point out that it is tempting to discuss our result (\ref{Ws3 in d=3})
in the context of holographic correspondence.
Indeed, the earlier works by \cite{Petkou:2003zz,Sezgin:2003pt}
(see also \cite{Giombi:2009wh} for an extensive discussion of the holographic three-point functions)
argue that to leading order in the large $N$
expansion, the higher-spin theory in $AdS_4$ bulk implies that in $d=3$ 
\begin{equation}
\label{sss general to leading in 1/N}
\langle sss\rangle = 0 + {\cal O}(1/N^{3/2})\,.
\end{equation}
This equation is in full agreement with the field theory
dual, which in fact has been known
since \cite{Petkou:1995vu,Petkou:1994ad}, as we reviewed above in this section.

The new result (\ref{Ws3 in d=3})
for the next-to-leading order $\langle sss\rangle$
raises a natural question whether the holographic work
of \cite{Petkou:2003zz,Sezgin:2003pt}
extends to order  ${\cal O}(1/N^{3/2})$.
Notice that \cite{Petkou:2003zz} shows that 
(\ref{sss general to leading in 1/N}) is valid for the $AdS_4$ dual of \textit{any} $O(N)$ CFT in $d=3$.
In particular, no assumption was made that the $AdS_4$ bulk is populated by the higher-spin fields
of Vasiliev's theory \cite{Petkou:2003zz}.
In fact, it may happen that (\ref{Ws3 in d=3}) holds for a large class of $O(N)$ CFTs in $d=3$. Holography is a possible tool to check this conjecture.
In particular, one can try to extend the holographic argument of \cite{Petkou:2003zz} to the next-to-leading order in the $1/N$ expansion.
A possible outcome $\langle sss\rangle = 0 + {\cal O}(1/N^{5/2})$ of a generic bulk calculation
would be a strong evidence in favor of the holographic duality \cite{Klebanov:2002ja},
and universality of the $W_{s^3}$ for a broad class of $O(N)$ CFTs in general dimension.

\section{Discussion}
\label{sec:discussion}

In this work we explore the $O(N)$ critical vector model with quartic interaction in $2\leq d \leq 6$ dimensions. In higher dimensions this model provides a nice illustration of the asymptotically safe quantum field theory, whereas in lower dimensions it is closely related to realistic systems such as the critical Ising model in 3D. While it is difficult to prove the existence of the fixed point in full generality, perturbative approach, $\epsilon$- and large-$N$ expansions confirm they exist in this model. Our calculations encompass the next-to-leading order analysis in the $1/N$ expansion and extend previously known results in a few ways.  

We derive and perform consistency checks that provide an additional evidence for the existence of a non-trivial fixed point. Moreover, we continue non-perturbative studies of the emergent conformal field theory and use conformal techniques to calculate a new CFT data associated with the three-point functions of the fundamental scalar and Hubbard-Stratonovich fields. This helps to expand our understanding of a CFT describing critical $\phi^4$ model in general dimension. Along the way  we evaluate a number of conformal multi-loop diagrams up to and including 4 loops in general $d$. These diagrams are generic and have value in themselves since they are not restricted to the critical $\phi^4$ model, but rather inherent to various CFTs.  

In \cite{Fei:2014yja,Fei:2014xta} an alternative description of the critical $O(N)$ model in terms of $N+1$ massless scalars with cubic interactions was proposed. It was explicitly shown that the scaling dimensions of various operators within the alternative description match the known results for the critical $O(N)$ theory. While our findings confirm the observations made in \cite{Fei:2014yja,Fei:2014xta}, we find an additional agreement between the next-to-leading
coefficients of the $\langle sss\rangle$ three-point functions, previously unobserved in the literature.

 In fact, it can be shown that all higher order correlation functions $\langle s(x_1)\cdots s(x_n)\rangle$, $n=4,5,\dots$ in both models match to leading order in the $1/N$ and $\epsilon$ expansion. In general, these correlators are entirely fixed by the 1PI vertices of the effective action. To leading order in the $1/N$ expansion these vertices are given by a single $\phi$-loop with $s$-legs attached to it. In the $\phi^4$ theory they have the following form
 \begin{equation}
 \label{n point function in phi4}
   \Gamma_{\phi^4}^\text{1PI}(x_1,\ldots x_n)
   =\,N\,C_\phi^n\,C_s^{n/2}\, \left(-\frac{2}{\sqrt{N}}\right)^n\,
 \int \prod_{i=1}^n d^dx_i ~ \mathcal{I} (x_1,\dots, x_n) s(x_1)\cdots s(x_n)\,,
 \end{equation}
 where $\mathcal{I} (x_1,\dots, x_n)$ is a space dependent structure representing the internal $\phi$-loop, and the external $s$-legs are normalized according to  (\ref{field normalizations}).
 In the cubic model the same vertex equals to
 \begin{equation}
   \label{n point function in cubic}
   \Gamma_\text{cubic}^\text{1PI}(x_1,\ldots x_n) =N\,C_\phi^{3n/2}\, (-g_1)^n\,
  \int \prod_{i=1}^n d^dx_i ~ \mathcal{I} (x_1,\dots, x_n)s(x_1)\cdots s(x_n)\,,
 \end{equation}
 where the fixed point value of the $\phi\phi s$ coupling constant is given by \cite{Fei:2014yja}
 \begin{equation}
 g_1 = \sqrt{\frac{6\epsilon (4\pi)^3}{N}}\left(1+\mathcal{O}\left(\frac{1}{N},\;\epsilon\right)\right)~.
 \end{equation}
In (\ref{n point function in cubic}) we took into account that the $s$ field in the cubic model is canonically
 normalized, and therefore the amplitude of its propagator in position space equals $C_\phi$.
 Note that (\ref{n point function in phi4}) and
 (\ref{n point function in cubic}) share the same structure $ \mathcal{I} (x_1,\dots, x_n)$ because the 1PI diagrams are identical up to an overall constant prefactor, which is explicitly written down in both cases. It then remains to show that these constants are identical. Indeed,
 \begin{equation}
 \left(\sqrt{N} \, \frac{g_1}{2}\right)^2\,\frac{C_\phi}{C_s} = 1 +\mathcal{O}\left(\frac{1}{N},\;\epsilon\right)\,.
 \end{equation}

The above matches support the equivalence suggested in\cite{Fei:2014yja,Fei:2014xta} between the critical $\phi^4$ vector model and the critical cubic model in $d=6-\epsilon$ dimensions.
Based on these findings we propose that the new result for $W_{s^3}$, which did not appear in the literature before our work,
 is in fact universally applicable to a large class of $O(N)$ CFTs. Therefore we interpret
the match between the $W_{s^3}$ coefficients of the critical $\phi^4$
 model and the critical cubic model in $d=6-\epsilon$ dimensions as a particular manifestation
 of this universality. Moreover, we suspect that it might be possible to systematically derive additional universal relations, at least in the $1/N$ expansion, thereby replacing the equivalence of  \cite{Fei:2014yja,Fei:2014xta}
 by a universally valid bootstrap statement.\footnote{Another reason to believe in the universality of the critical $O(N)$ vector models in general $d$ is due to the observation related to the behavior
 of $W_{s^3}$ coefficient
 when extrapolated to $d=8$. While the vector model in $d=8$
 is manifestly non-unitary, since the scaling dimension of the operator $s$ is below
 the unitarity bound, we can nevertheless formally compare our result $NW_{s^3}(d=8) = -950$ with a similar calculation in the exotic critical model in $8-\epsilon$ dimensions \cite{Gracey:2015xmw}. We find a precise match with $NW_{s^3}$ calculated in \cite{Gracey:2015xmw}.
 We are grateful to Simone Giombi, Igor Klebanov and Gregory Tarnopolsky for letting us know about
 \cite{Gracey:2015xmw} and suggesting to compare the results.}

Furthermore, we notice that
 the critical $O(N)$ model in higher dimensions does have certain unphysical features. The well-known $\epsilon$-expansion shows that the coupling constant at the Wilson-Fisher fixed point is negative in $4+\epsilon$ dimensions \cite{Weinberg:1976xy}. This means that the potential is unbounded from below for large values of the field. However, perturbative calculations in $1/N$ do not reveal any sign of instability at the level of the correlation functions. Thus, for instance, unitarity bounds are satisfied. Of course, this argument only shows that the vacuum state of the theory is metastable if the instability observed within $\epsilon$-expansion persists in the $\epsilon\to 1$ limit.

Indeed, in the recent work \cite{Giombi:2019upv} the authors provide a non-perturbative argument in favour of instability of the model. Their analysis rests on the observation that the path integral in the large-$N$ limit is entirely dominated by the saddle point. The corresponding saddle point equation is Weyl invariant at the fixed point, and therefore admits a family of solutions parametrized by size and location. This type of large-$N$ instantons was previously observed in the critical $\phi^6$ model in three dimensions \cite{Smolkin:2012er}. In particular, it was argued that the instantons and associated instability of the critical $\phi^6$ model can be used as a toy model towards holographic resolution of the singularity and conformal factor problems in quantum cosmology. It would be interesting to explore these aspects in the context of critical $\phi^4$ model. 

Remarkably, the critical $O(N)$ vector models exhibit peculiar behaviour when coupled to a thermal bath \cite{Chai:2020zgq}. We hope that certain results presented in this paper might be of help towards understanding this behaviour in lower dimensions. Progress in this direction will be reported elsewhere \cite{prog}.

\section*{Acknowledgements} \noindent We thank Noam Chai, Soumangsu Chakraborty, Johan Henriksson, Zohar Komargodski and Anastasios Petkou for helpful discussions and correspondence. We would like to express our special thanks of gratitude to Simone Giombi, Igor Klebanov and Gregory Tarnopolsky for numerous comments and stimulating correspondence which helped us to improve our results and their presentation.  This work is partially supported by the Binational Science Foundation (grant No. 2016186), the Israeli Science Foundation Center of Excellence (grant No. 2289/18) and by the Quantum Universe I-CORE program of the Israel Planning and Budgeting Committee (grant No. 1937/12).

\appendix

\section{Callan-Symanzik equations}
\label{appendix:callan-symanzik equation}

In this appendix we use the Callan-Symanzik equation to derive various relations
between the counterterms $\delta_\phi$, $\hat\delta _s$, $\hat \delta_4$, and the anomalous dimensions
$\gamma_\phi$, $\gamma_s$. In particular we establish the identity (\ref{delta s hat delta s result})
satisfied by the counterterms $\delta_s$ and $\hat\delta _s$. The calculation is carried out for the model (\ref{renormalized action}) with mass set to zero $m=0$. Depending on the dimension $d$ the model is assumed to sit either at the UV or IR fixed point. 

Let us start with the two-point function for the renormalized field $\phi$,
\begin{equation}
\langle \tilde\phi(p)\tilde\phi(q)\rangle
= (2\pi) ^ d \delta(p+q) \left( \frac{1}{p^2} + \textrm{loop corrections} + \frac{-p^2\,\delta_\phi}{(p^2)^2} \right)\,.
\end{equation}
It satisfies the Callan-Symanzik equation
\begin{equation}
\left(\mu\frac{\partial}{\partial\mu} + 2\gamma _ \phi\right) \langle \tilde\phi(p)\tilde\phi(q)\rangle = 0\,,
\end{equation}
which leads to 
\begin{equation}
\label{gamma phi in terms of delta phi}
\gamma _ \phi = \frac{1}{2}\mu\frac{\partial}{\partial\mu}\,\delta_\phi + \mathcal{O}(1/N^2)\,.
\end{equation}
Similarly, the Callan-Symanzik equation for the Hubbard-Stratonovich field reads
\begin{equation}
\label{ss Callan-Symanzik  equation}
\left(\mu\frac{\partial}{\partial\mu} + 2\gamma _ s\right) \langle \tilde s(p)\tilde s(q)\rangle = 0\,,
\end{equation}
Here we have 
\begin{equation}
\label{ss propagator for CS equation}
\langle \tilde s(p)\tilde s(q)\rangle
= (2\pi) ^ d \delta(p+q) \left( - \frac{1}{2B(p)} + \textrm{loop corrections} + \frac{\hat \delta_s}{2B(p)} \right)\,,
\end{equation}
where $B$ is given by (\ref{B in general d}).
At the order ${\cal O}(1/N)$ the loop corrections to the $\langle \tilde s\tilde s\rangle$
propagator in (\ref{ss propagator for CS equation}) contain dependence on the renormalization scale $\mu$, due to contributions
from the $\phi\phi s$ interaction vertex counterterm, and the $\phi$ field strength
renormalization counterterm. Denoting the associated
terms with ${\cal L}_s(\mu)$ and using (\ref{ss Callan-Symanzik equation})
we obtain
\begin{equation}
\label{gamma s in terms of hat delta s}
\gamma _ s = \frac{1}{2}\mu\frac{\partial}{\partial\mu}\,(\hat \delta_s +
2B{\cal L}_s)+ \mathcal{O}(1/N^2)\,.
\end{equation}
Finally, for the three-point function
\begin{align}
\langle \tilde \phi (p _ 1) \tilde \phi (p _ 2) \tilde s ( q ) \rangle
&=(2\pi) ^ d \delta (p _ 1 + p _ 2 + q)\,\frac{1}{p_1^2}\frac{1}{p_2^2}\frac{-1}{2B(q)} \\
&\times \left( -\frac{2}{\sqrt{N}}  + \textrm{loop corrections} -\frac{2}{\sqrt{N}} \frac{\hat \delta _4}{\sqrt{\tilde g_4}}
+(-2\delta _\phi - \hat \delta _s)\left(-\frac{2}{\sqrt{N}}\right)\right)\,,\notag
\end{align}
the Callan-Symanzik equation at the fixed point takes the form
\begin{equation}
\left(\mu\frac{\partial}{\partial\mu} + 2\gamma _ \phi + \gamma_s
\right)\langle \tilde \phi (p _ 1) \tilde \phi (p _ 2) \tilde s ( q ) \rangle
 = 0.
\end{equation}
As a result, we obtain 
\begin{equation}
\mu\frac{\partial}{\partial\mu}\left(\delta_s
- \hat \delta _s  +2 B  {\cal L}_s \right)= 0+\mathcal{O}(1/N^2)\,,
\end{equation}
where we substituted (\ref{gamma phi in terms of delta phi}), (\ref{gamma s in terms of hat delta s}) and used  (\ref{delta s in terms of auxiliaries}).

\section{Calculation of the trapezoid diagram}
\label{appendix: calculation of v3}

In this appendix we calculate the trapezoid diagram $\tilde v_3(\delta)$
defined by (\ref{v3 in terms of tilde v3}). For completeness we reproduce here the defining diagram 
\begin{center}
  \begin{picture}(165,90) (75,-5)
    \SetWidth{1.0}
    \SetColor{Black}
    \Line[](32,6)(192,6)
    \Line[](192,6)(152,70)
    \Line[](72,70)(32,6)
    \Line[](112,6)(72,74)
    \Line[](72,70)(152,70)
    \Line[](152,70)(112,6)
    \Vertex(72,70){4}
    \Vertex(152,70){4}
    \Vertex(32,6){2}
    \Vertex(112,6){4}
    \Vertex(192,6){2}
    \Text(17,-5)[lb]{$x_2$}
    \Text(198,-5)[lb]{$x_3$}
    \Text(68,-5)[lb]{\scalebox{0.6}{$2$}}
    \Text(183,35)[lb]{\scalebox{0.6}{$2-\delta$}}
    \Text(110,76)[lb]{\scalebox{0.6}{$2$}}
    \Text(82,35)[lb]{\scalebox{0.6}{$2$}}
    \Text(137,35)[lb]{\scalebox{0.6}{$2$}}
    \Text(145,-5)[lb]{\scalebox{0.6}{$2d-6$}}
    \Text(10,35)[lb]{\scalebox{0.6}{$2d-6+2\delta$}}
    \Text(220,30)[lb]{$=\frac{\tilde v_3(\delta )}{|x_{23}|^{d-2+\delta}}$}
  \end{picture}
\end{center}
It turns out that this diagram is quite ubiquitous, because various conformal graphs can be reduced to the trapezoid form through the use of conformal techniques,  see for instance equations (4.345) and (4.346) in \cite{Vasilev:2004yr}. 
Although we did not find an explicit analytic expression for this diagram in the literature, the diagrammatic identity (4.345)  in \cite{Vasilev:2004yr} suggests the following relation
$\tilde v_3(0) = \frac{U\left(1,2,d-3\right)}{U\left(\frac{d}{2}-1,\frac{d}{2}-1,2\right)^2}\times\lim_{\lambda'\to 0} (4.346)$. The direct calculation of $\tilde v_3(0)$ in this Appendix matches this result.

To calculate the trapezoid graph we use the following relation obtained by integration by parts \cite{Chetyrkin:1981qh}, see also \cite{Kazakov:1983ns,Kazakov:1984bw}
\begin{center}
  \begin{picture}(250,192) (31,-130)
    \SetWidth{1.0}
    \SetColor{Black}
    \Line[](64,6)(32,-26)
    \Line[](64,6)(96,-26)
    \Line[](64,6)(64,54)
    \Vertex(64,6){4}
    \Text(50,27)[lb]{\scalebox{0.6}{$2\alpha_1$}}
    \Text(47,-20)[lb]{\scalebox{0.6}{$2\alpha_2$}}
    \Text(72,-20)[lb]{\scalebox{0.6}{$2\alpha_3$}}
    \Text(-40,3)[lb]{\scalebox{0.6}{$(d-2\alpha_1-\alpha_2-\alpha_3)\times$}}
    \Line[](174,6)(142,-26)
    \Line[](174,6)(206,-26)
    \Line[](174,6)(174,54)
    \Vertex(174,6){4}
    \Text(140,27)[lb]{\scalebox{0.6}{$2(\alpha_1-1)$}}
    \Text(117,-20)[lb]{\scalebox{0.6}{$2(\alpha_2+1)$}}
    \Text(185,-20)[lb]{\scalebox{0.6}{$2\alpha_3$}}
    \Text(120,3)[lb]{\scalebox{0.6}{$=\;\;\alpha_2\times\;\;$}}
    \Line[](284,6)(252,-26)
    \Line[](284,6)(316,-26)
    \Line[](284,6)(284,54)
    \Line[](252,-26)(284,54)
    \Vertex(284,6){4}
    \Text(290,27)[lb]{\scalebox{0.6}{$2\alpha_1$}}
    \Text(267,-20)[lb]{\scalebox{0.6}{$2(\alpha_2+1)$}}
    \Text(315,-20)[lb]{\scalebox{0.6}{$2\alpha_3$}}
    \Text(230,3)[lb]{\scalebox{0.6}{$-\;\;\alpha_2\times\;\;$}}
    \Text(255,20)[lb]{\scalebox{0.6}{$-2$}}
    \Line[](174,-94)(142,-126)
    \Line[](174,-94)(206,-126)
    \Line[](174,-94)(174,-46)
    \Vertex(174,-94){4}
    \Text(140,-73)[lb]{\scalebox{0.6}{$2(\alpha_1-1)$}}
    \Text(137,-120)[lb]{\scalebox{0.6}{$2\alpha_2$}}
    \Text(165,-120)[lb]{\scalebox{0.6}{$2(\alpha_3+1)$}}
    \Text(120,-97)[lb]{\scalebox{0.6}{$+\;\;\alpha_3\times\;\;$}}
    \Line[](284,-94)(252,-126)
    \Line[](284,-94)(316,-126)
    \Line[](284,-94)(284,-46)
    \Line[](316,-126)(284,-46)
    \Vertex(284,-94){4}
    \Text(270,-73)[lb]{\scalebox{0.6}{$2\alpha_1$}}
    \Text(247,-120)[lb]{\scalebox{0.6}{$2\alpha_2$}}
    \Text(275,-120)[lb]{\scalebox{0.6}{$2(\alpha_3+1)$}}
    \Text(230,-97)[lb]{\scalebox{0.6}{$-\;\;\alpha_3\times\;\;$}}
    \Text(300,-80)[lb]{\scalebox{0.6}{$-2$}}
  \end{picture}
\end{center}
We apply this relation to the top left vertex of the $\tilde v_3$ diagram
(with $\alpha_1=1$, $\alpha_2=d-3+\delta$, $\alpha_3=1$), getting a sum of four terms.
Then we use the propagator merging relation (\ref{propagator splitting}) to integrate over one of the
vertices in the first three of those terms. As a result, the first three terms acquire the form of a self-energy diagram,
and the fourth terms acquires the form of a self-energy diagram with a line attached to the left-hand vertex.
These self-energy diagrams can be brought to a similar form by an inversion transformation around
the right-hand vertex.
In the end we obtain
\begin{align}
\label{tilde v3 expression}
\tilde v_3(\delta) &= -\frac{1}{\delta}\,\left[
\left(U(d-3+\delta,2,1-\delta)+(d-3+\delta)U(d-2+\delta,1,1-\delta)\right)\,d_1
\right.\\
&-\left.U\left(2,1-\frac{\delta}{2},d-3+\frac{\delta}{2}\right)\,d_2
-(d-3+\delta)\, U\left(d-2+\delta,1-\frac{\delta}{2},1-\frac{\delta}{2}\right)\,d_3\right]\,,\notag
\end{align}
where we denoted
\begin{align}
d_1 &= F\left(\frac{d}{2}-1+\delta, \, \frac{d}{2} - 1 - \frac{\delta}{2}\right)\,,\\
d_2 &= F\left(d-3+\delta, \, \frac{d}{2}-1-\frac{\delta}{2}\right)\,,\\
d_3 &= F\left(1,\, d-3+\frac{\delta}{2}\right)
\end{align}
for the function $F(\alpha,\beta)$ defined by the self-energy diagram
\begin{center}
  \begin{picture}(228,130) (80,-15)
    \SetWidth{1.0}
    \SetColor{Black}
    \Line[](48,50)(128,98)
    \Line[](128,98)(208,50)
    \Line[](48,50)(128,2)
    \Line[](128,2)(208,50)
    \Line[](128,98)(128,2)
    \Vertex(128,98){4}
    \Vertex(128,2){4}
    \Text(80,18)[lb]{\scalebox{0.8}{$2$}}
    \Text(176,18)[lb]{\scalebox{0.8}{$2$}}
    \Text(135,50)[lb]{\scalebox{0.8}{$2$}}
    \Text(72,82)[lb]{\scalebox{0.8}{$2\alpha$}}
    \Text(170,82)[lb]{\scalebox{0.8}{$2\beta$}}
    \Text(32,50)[lb]{\scalebox{0.8}{$x_1$}}
    \Text(217,50)[lb]{\scalebox{0.8}{$x_2$}}
    \Vertex(48,50){2}
    \Vertex(208,50){2}
    \Text(250,42)[lb]{$=\frac{F(\alpha,\beta)}{|x_{12}|^{2(3+\alpha+\beta-d)}}$}
  \end{picture}
\end{center}
 To calculate this diagram we can again use the integration by parts relation for
 the top vertex, with $\alpha_1=1$, $\alpha_2=\alpha$, $\alpha_3=\beta$,
 after which all the remaining integrals can be straightforwardly taken using
 (\ref{propagator splitting}).
 As a result, we obtain
\begin{align}
F(\alpha,\beta) &= \frac{U(1,1,d-2)}{d-2-\alpha-\beta}\,
\left[\alpha \,\left(U(\alpha+1,\beta,d-\alpha-\beta-1)
-U\left(\alpha+1,\beta+2-\frac{d}{2},\frac{3d}{2}-\alpha-\beta-3\right)\right)
\right.\notag\\
&+\left.\beta \,\left(U(\alpha,\beta+1,d-\alpha-\beta-1)
-U\left(\alpha+2-\frac{d}{2},\beta+1,\frac{3d}{2}-\alpha-\beta-3\right)\right)\right]\,.
\end{align}
This diagram is dual by Fourier transform to the ChT diagram given by eq. (16) in \cite{Vasiliev:1981dg},
as can be checked explicitly 
\begin{equation}
F(\alpha,\beta) = \frac{A(1)^3A(\alpha)A(\beta)}{A(3+\alpha+\beta-d)}\,
\textrm{ChT}\left(\frac{d}{2}-\alpha,\frac{d}{2}-\beta\right)\,,
\end{equation}
where
\begin{align}
\label{ChT expression}
\textrm{ChT}(\alpha,\beta)&=
\frac{\pi ^d A(d-2) \left(\frac{A(2-\alpha ) A(\alpha )}{(1-\beta ) (\alpha +\beta -2)}+\frac{A(\alpha +\beta -1) A(-\alpha -\beta +3)}{(\alpha -1) (\beta -1)}+\frac{A(2-\beta ) A(\beta )}{(1-\alpha ) (\alpha +\beta -2)}\right)}{\Gamma \left(\frac{d}{2}-1\right)}\,,\\
A(\Delta) &= \frac{\Gamma\left(\frac{d}{2}-\Delta\right)}{\Gamma(\Delta)}\,.
\label{A of Delta definition}
\end{align}
As a result, from (\ref{tilde v3 expression}) we obtain
\begin{align}
\tilde v_3(0) &=\frac{-2^{4-d} \pi ^{\frac{3 d}{2}+
\frac{5}{2}} }{ (d-4)\sin^2\big({\pi d\over 2}\big) \Gamma \left(\frac{d-3}{2}\right) \Gamma (d-2)}\notag\\
&\times\left( \psi ^{(1)}(d-3) -6 \psi ^{(1)}\Big(\frac{d}{2}-1\Big)+H_{d-4}\Big(H_{d-4}+2\pi\cot\big({\pi d\over 2}\big) \Big)  +{\pi^2\over 6} 
\right) 
\,.\label{tilde v3 answer}
\end{align}
 The $\epsilon$-expansion of  the trapezoid diagram in $d=4-2\epsilon$ dimensions was carried out in \cite{Kazakov:1983ns} (see eq. (19) therein) and \cite{Kazakov:1984bw} (see eq. (2.24) therein)  to ${\cal O}(\epsilon^2)$ order.\footnote{Conventions of these references are slightly different from ours. One would need to set $a_1=a_4=-2$, $a_2=a_3=a_5=a_6=a_7=0$, and keep in mind that those references put the label equal to a half of the exponent on top of the propagator lines. In conventions of \cite{Kazakov:1983ns,Kazakov:1984bw} there is also a factor of $\pi^{-\frac{d}{2}}$ in each vertex.} Hence, we can test our result (\ref{tilde v3 answer}) by substituting $d=4-2\epsilon$, expanding in $\epsilon$ and comparing with \cite{Kazakov:1983ns,Kazakov:1984bw}.\footnote{We thank Simone Giombi, Igor Klebanov and Gregory Tarnopolsky for the related discussion.} Performing the comparison we obtain a precise match to all available orders (\textit{i.e.}, up to ${\cal O}(\epsilon^2)$)
of the $\epsilon$-expansion.

\end{document}